\documentclass{aa}
\usepackage{graphics}
\def\hi{H\,{\sc i} }

\def\kmss{km~s$^{-1}$ }
\def\kms{km~s$^{-1}$}

\def\msol{${\rm M}_{\sun}$}
\def\smppc2{${\rm M}_{\sun} {\rm pc}^{-2}$}
\def\rcore{R_{\rm core}}
\def\kprime{K^{\prime}}
\def\vhmax{v_{\rm h}^{\rm max}}
\begin{document}
%\thesaurus{03(11.09.1 NGC 3992;  09.11.1 11.06.2 11.11.1 11.19.2)} 
\title{Dark and luminous matter in the NGC 3992 group of galaxies}
\subtitle{II. The dwarf companions UGC 6923, UGC 6940, UGC 6969, and
the Tully-Fisher relation}
\author{Roelof Bottema}
%\offprints{R. Bottema}
\institute{Kapteyn Astronomical Institute, P.O. Box 800, 
NL-9700 AV Groningen, The Netherlands, e-mail:robot@astro.rug.nl}
\date{Received date1; accepted date2}
\abstract{
Detailed neutral hydrogen observations have been obtained of
the large barred  spiral galaxy NGC 3992 and its three small companion
spiral galaxies, UGC 6923, UGC 6940, and UGC 6969. 
Contrary to the large galaxy, for the companions the \hi distribution
ends quite abruptly at the optical edges. Velocity fields have
been constructed from which rotation curves have been derived.
Assuming a reasonable M/L ratio, a decomposition
of these rotation curves generates nearly equal dark matter halos.
When comparing the position-velocity diagrams of the two 
brightest galaxies, UGC 6923 and UGC 6969, it is obvious
that the rotation curve of the latter has a shape closer to
solid body than the former, yet the same maximum rotational
level is reached. This is likely generated by the equal dark matter
halos in combination with UGC 6923 being
a factor five more luminous than UGC 6969 and so its luminous
matter gives a higher contribution to the rotation in
the inner regions. An NFW-CDM$\Lambda$ dark halo is consistent
with the observed rotation curve of UGC 6923 but not consistent
with the rotation curve of UGC 6969.
If the NGC 3992 group is part of the Ursa Major cluster, then
the I-band $M/L$ ratio of NGC 3992 has to be at least 1.35 times as large
as that of the average spiral galaxy in the cluster.
On the other hand, equal $M/L$ ratios can be achieved when
the NGC 3992 group is placed more than 3 Mpc behind the cluster.
Both possibilities can explain why NGC 3992 appears to be 0.43
magnitudes too faint for its rotation.
}
\maketitle
\section{Introduction}

Rotation curves derived from neutral hydrogen observations at
the outer regions of spiral galaxies unambiguously show that
substantial amounts of dark matter are required (Bosma 1978;
Begeman 1987, 1989). Any physically reasonable distribution
of this dark matter necessitates the presence of at least some of
that in the inner optical disc region, contributing in some degree
to the total rotation in that region. Unfortunately, from the
observed rotation curve and light distribution one cannot a priori
determine the ratio of dark to luminous matter (van Albada et al. 1985).
There are arguments, mainly theoretical, that the contribution of
the disc has to be maximized, leading to the so called maximum
disc hypothesis (van Albada \& Sancisi 1986;
Sellwood \& Moore 1999). On the other hand, observations of disc
stellar velocity dispersions (Bottema 1993, 1997) lead to the conclusion
that the disc contributes, on average, 63\% to the total rotation
at the position where the disc has its maximum rotation. This finding
is supported by a statistical analysis of rotation curve shapes in
relation to the compactness of discs (Courteau \& Rix 1999). 
If the rotation is not measured outside the optical disc,
the observed rotation curve can in most cases be explained by
the stellar components alone (Kalnajs 1983; Kent 1986).
This statement is valid for normal, medium and large galaxies.
However, for smaller less massive or for galaxies with lower
surface brightness, one does need dark matter already within
the optical disc (Salucci et al. 1991; de Blok \& McGaugh 1997).
This indicates that small galaxies hold clear clues as to the
size and shape of dark halos.

For a long time dark halos have been described by a pseudo
isothermal sphere (Carignan \& Freeman 1985). Such a halo 
has a constant density core and its rotation curve is characterized
by two parameters. Numerical calculations of structure formation
in a cold dark matter (CDM) dominated universe generate a different
dark halo (Navarro et al. 1996, 1997). It has a central cusp
where the density goes like $r^{-1}$. When a certain cosmology
is chosen the structural scale is related to the total mass.
Rotation curves are then characterized by only one parameter
and are as such less flexible than rotation curves of isothermal
halos. These halos are commonly referred to as NFW halos and seem
to be able to fit the observed rotation curves of normal
galaxies just as well (Navarro 1998). There is, however, quite
some debate whether NFW halos can also fit the rotation
curves of low surface brightness (hereafter LSB) galaxies.

At least when measured in kpc, rotation curves of LSB galaxies seem
to rise slower in the inner regions than those of normal
galaxies with the same luminosity (de Blok \& McGaugh 1997; 
Pickering et al. 1997). That implies that for LSB systems the stellar
disc can nowhere be dominant. If a disc is forced to its maximum
contribution that leads in general to an excessively large
mass-to-light ratio, while it is expected on the basis of colours
and of metallicity that LSB galaxies have rather low M/L ratios.
The establishment of slowly rising rotation curves of LSB galaxies
is based on \hi line observations.
On the other hand, additional observations in the H$\alpha$ line 
by Swaters et al. (2000) showed that in some cases beam smearing
may have caused the \hi rotation curves to be shallower than in
reality. NFW halos cannot be reconciled with slowly rising,
solid body rotation curves because for an $r^{-1}$ density profile
one has an $r^{1/2}$ rotation curve. For that reason it is
rather important to establish the exact rotational behaviour
of LSB galaxies in their inner regions. 
The structure formation calculations
generate NFW halos on all linear scales. Therefore, if there is
a class of objects for which NFW halos do not apply the CDM paradigm
might be in serious trouble. In a recent paper (de Blok et al. 2001)
radial density profiles as generated by H$\alpha$ rotation curves
are presented for a number of LSB systems. In at least five
out of $\sim$ 40 cases the profiles cannot be reconciled with
NFW halos.

Comparing the normal galaxies to the LSB systems in the Ursa Major
cluster, Tully \& Verheijen (1997) conclude that certainly
the LSB galaxies must be sub maximal. The NGC 3992 group is part
of the Ursa Major cluster and the three companions of NGC 3992
are small LSB galaxies. Consequently, observations and analysis of the
rotation curves of these companions is of interest.
Not only is it important to make a comparison with the other
galaxies in the cluster, but the derived rotation curves may also
shed extra light on the universality of CDM generated mass structures.  

The NGC 3992 group consists of the large barred spiral galaxy
NGC 3992 and its three small companion galaxies UGC 6923, UGC 6940,
and UGC 6969. The group had already been observed in \hi by
Gottesman et al. (1984) and later by Verheijen (1997)
and by Verheijen \& Sancisi (2001). Both these
observations suffer from a limited resolution or limited S/N level.
In order to derive a detailed rotation curve of a large barred spiral,
NGC 3992 was observed again with the Westerbork Synthesis Radio
Telescope. The results and analysis of these observations
are presented by Bottema \& Verheijen (2002, hereafter
Paper~I). Within the same field there are the \hi structures of
the companions. Because the total amount of material appeared to be
quite comprehensive and because the subjects of barred and
dwarf galaxies are relatively unrelated, it was decided to present
matters in two papers. A detailed description of the observations
and data handling has already been given in Paper~I; here only the observing
parameters are summarized in Table~1.

For the UMa cluster a Tully-Fisher relation (Tully \& Fisher 1977) has been
constructed by Verheijen (1997), which shows two
anomalies for the NGC 3992 group. Firstly, the massive galaxy
NGC 3992 is too faint for its maximum rotation. Related to this,
the derived mass-to-light ratio of that galaxy is nearly a factor
two larger than that of other spirals. Secondly, the two companions
UGC 6923 and UGC 6969 have the same maximum rotation yet differ
by a factor of five in total luminosity. That represents a clear source
of scatter in the Tully-Fisher relation. To investigate these
matters, luminosities, colours, and M/L ratios of the group members
will be compared in the present paper. For convenience a listing
of the main parameters of the galaxies is given in Table~2.

%Fig1
\begin{figure}
\resizebox{\hsize}{!}{\includegraphics{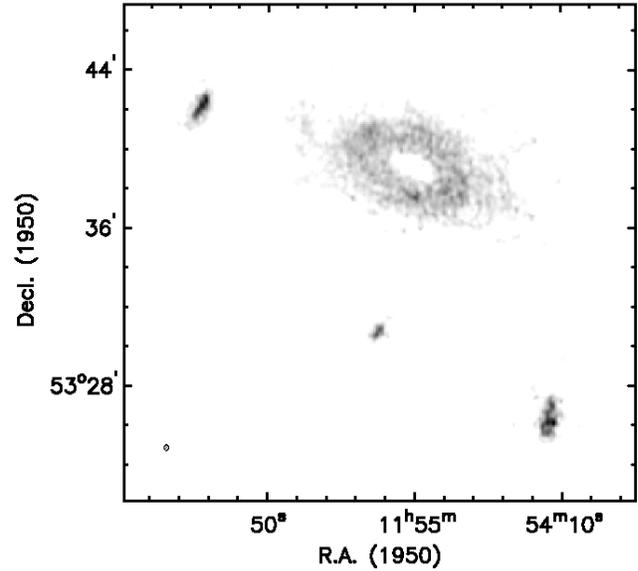}}
\caption[]{Greyscale image showing the full resolution
total \hi map of NGC 3992 and its surroundings.
From top left to bottom right the three companions,
UGC 6969, 6940, and 6923 are clearly visible;
their \hi column densities are larger than that of
the main galaxy. Note the central \hi hole of
NGC 3992, at the region of the bar.
The beam is indicated in the lower left, the greyscale
is linear from 0.2~10$^{20}$ to 34.8~10$^{20}$ H-atoms cm$^{-2}$.}
\end{figure}

The distance to the UMa cluster as a whole and to NGC 3992 
in particular has not yet been determined precisely.
Sakai et al. (2000) give a distance of 20.7 $\pm$ 3.2 Mpc
following from a Tully-Fisher analysis using the cepheid distances
to local galaxies. On the other hand, for a similar analysis,
Tully \& Pierce (2000) derive a distance to the UMa cluster of
18.6 Mpc, probably with the same error as that of Sakai et al.
In a recent re-evaluation of the HST distance scale project
(Freedman et al. 2001) the distances to the local calibrator
galaxies have decreased by $\sim$ 5\% and consequently the distances
to UMa of 20.7 and 18.6 Mpc should also be decreased by that amount. 
As for now, a distance of 18.6 Mpc seems reasonable and has been
adopted in the present paper. This distance differs, however,
from the 15.5 Mpc used in earlier studies of the UMa cluster
by Tully et al. (1996) and by Verheijen (1997).

\section{Total \hi}

To have an overview of the \hi distribution in the group,
the total \hi map at full resolution is displayed in Fig. 1.
Immediately obvious is the central hole in the gas distribution
of NGC 3992 at exactly the position of the bar. Furthermore, 
NGC 3992 has a faint gas extension outside its stellar disc.
This is contrary to the companions where the
\hi distribution ends abruptly at the
optical edge of the galaxy. 
A possible explanation for this is stripping of the gas from
the companions when these have passed by, or interacted with NGC 3992.
For the companions no effort has been made to construct
an \hi distribution at lower resolution. That would in practice
mainly result in a smearing of the gas somewhat over the edge, thereby
dominating the additional gas that might be there. Moreover,
when determining the gas kinematics for the companions the
additional sensitivity at lower resolution does not outweigh
the benefits of the higher resolution.

The position and amount of \hi gas has been determined the same
way as for the main galaxy; using the conditional transfer method
for the full resolution data. For all galaxies the same smoothing
and filtering parameters have been used (see Paper~I). Addition of
the signal in each channel map at the position of the galaxies gives
the \hi profiles. These were already displayed in Fig. 4 of Paper I,
together with that
of NGC 3992. All the signal of the companions falls within the \hi
profile of the main galaxy. Integration of the data cube 
along the velocity direction
results in the total \hi maps. These are displayed on the optical image
in Figs. 2, 3, and 4 for respectively UGC 6923, 6940, and 6969.
For the three galaxies the neutral hydrogen gas generally coincides
with the optical galaxies; there are no large displacements or
attached filaments. In the case of UGC 6923 (Fig. 2) a slight correspondence
of the gas with optical arms is visible. For UGC 6969 (Fig. 4) there
is somewhat more gas on the North West side compared to the South East
side. The small protrusion on the South East side has a velocity
deviating by 40 \kmss from the general velocity field. It has
approximately a radial velocity halfway between the value of
the general field and the systemic velocity (see Fig. 9, top panel). 
It might be a gas cloud
accreting on the galaxy or in the process of being tidally stripped.
Anyway, it has not been taken into account for the determination of
the rotation curve. The smallest galaxy (UGC 6940) is so moderately
resolved that not much can be said about the \hi gas in relation to
the optical, except that in general both coincide.

%Fig2
\begin{figure}
\resizebox{7.5truecm}{!}{\includegraphics{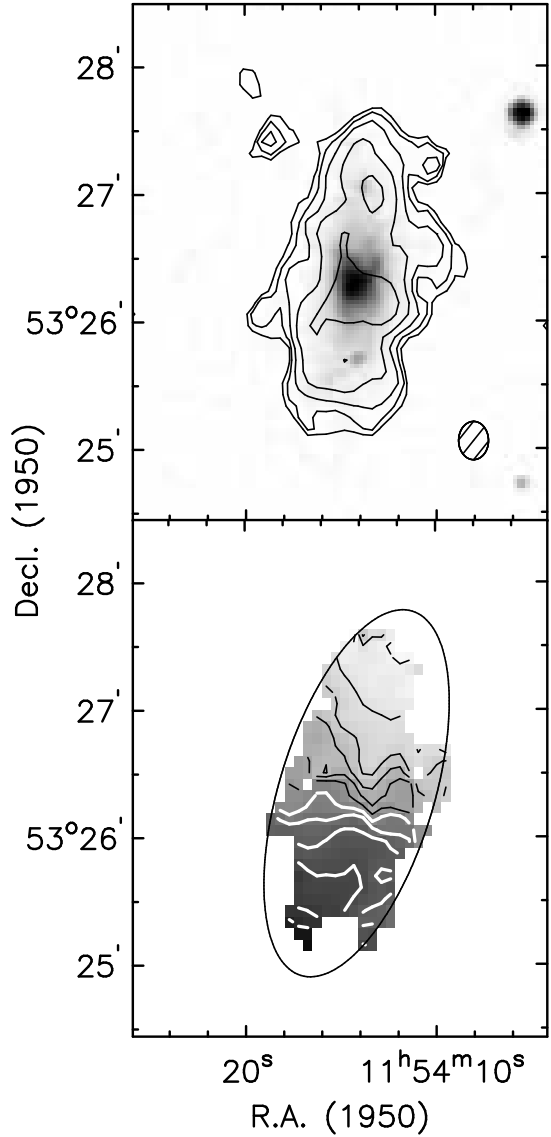}}
\caption[]{{\it Top:} Full resolution total \hi map of UGC 6923
superposed on the optical image. Contour levels increase
by a factor of two from 1.39~10$^{20}$ to 44.3~10$^{20}$ H-atoms cm$^{-2}$.
{\it Bottom: } Full resolution velocity field of UGC 6923.
The first black contour next to the white contours is at
the systemic velocity of 1066 \kms. Contours differ
by 15 \kmss and increase from bottom to top.
The ellipse indicates the position of the outermost
tilted ring for which a rotational value has been determined.}
\end{figure}

%Fig3
\begin{figure}
\resizebox{\hsize}{!}{\includegraphics{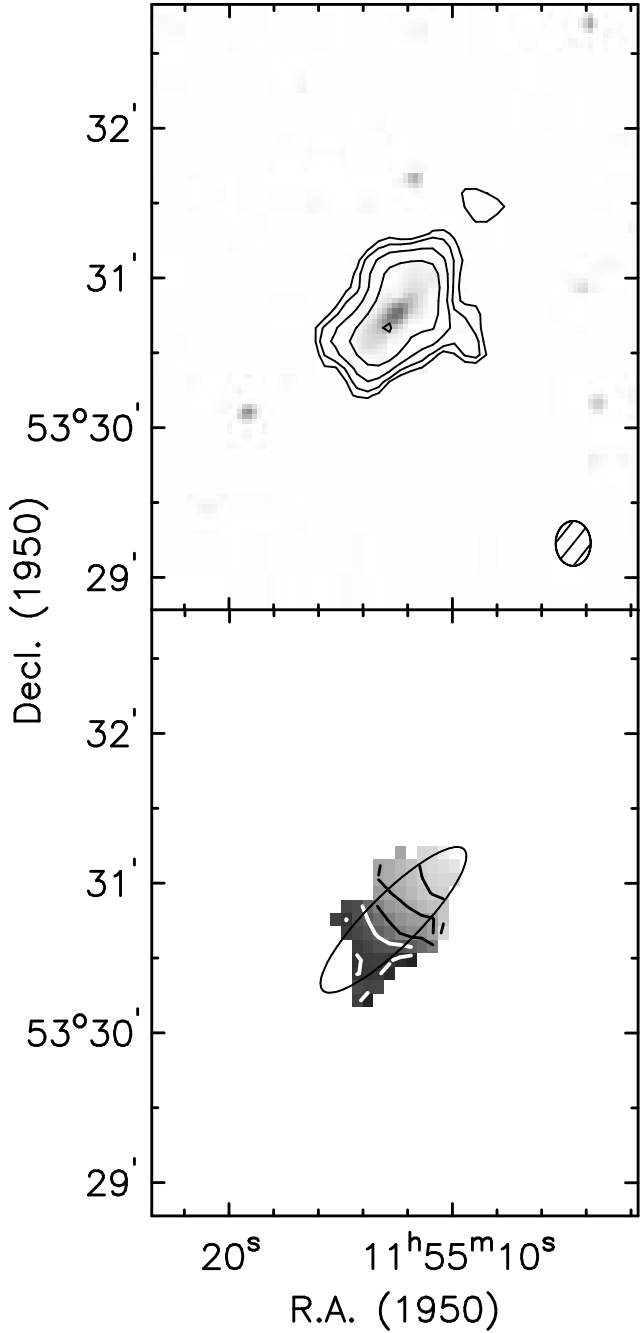}}
\caption[]{As for Fig. 2, but now for UGC 6940 having a
systemic velocity of 1107 \kms.}
\end{figure}

%Fig4
\begin{figure}
\resizebox{\hsize}{!}{\includegraphics{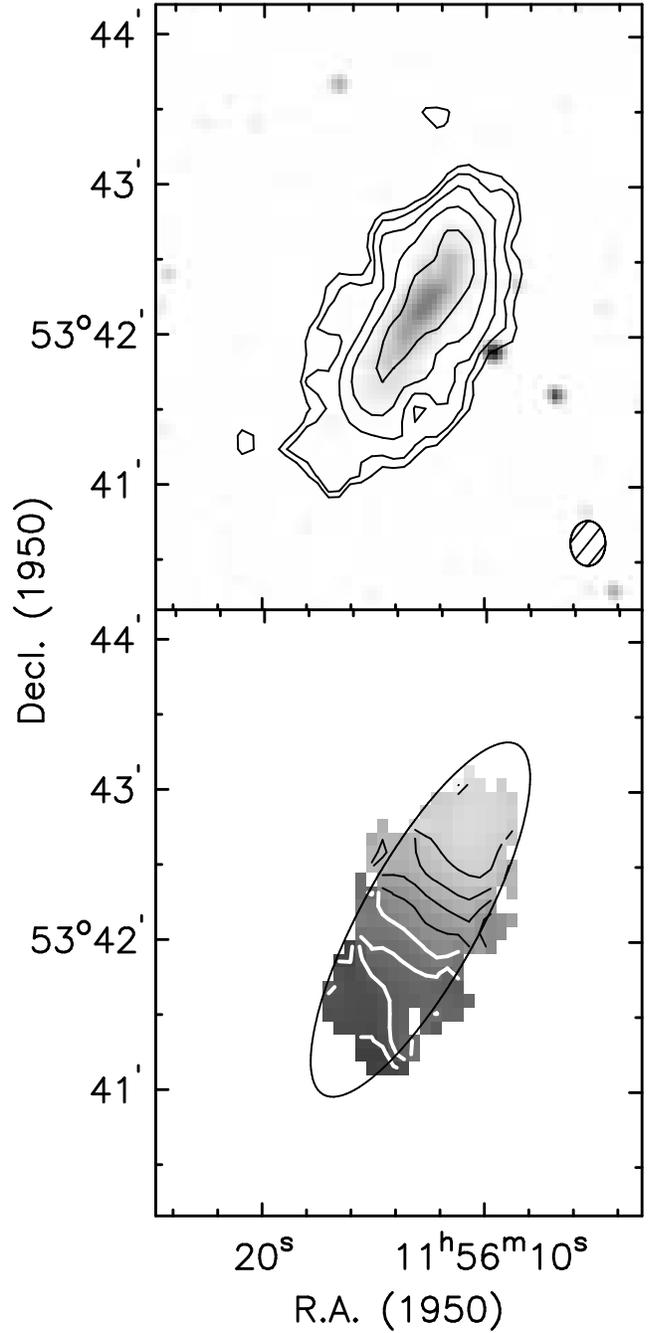}}
\caption[]{As for Fig. 2, but now for UGC 6969 having a
systemic velocity of 1114 \kms.}
\end{figure}

Integration over velocity and space gives the total \hi flux, which
amounts to 7.82, 1.92, and 5.43 Jy \kmss for respectively UGC 6923,
6940, and 6969. For a distance of 18.6 Mpc this translates in total
\hi masses of respectively 6.38~10$^8$, 1.57~10$^8$, and 4.44~10$^8$
\msol.

The surface density as a function of radius has been constructed
the same way as for the main galaxy. Elliptical annuli with orientations
given by the rotation curve fit were projected on the surface brightness
map. The emission was averaged over these annuli and scaled such that
integration of the radial profile gives the same total \hi mass
as found above. The resulting radial density profile is shown in
Fig. 5 for all three companions.
For UGC 6923 and UGC 6969 there is only a small amount of beam
smearing and the derived radial density profile is a
good representation of the actual gas distribution. This is not
valid for UGC 6940; in that case the profile given in Fig. 5 may
deviate somewhat from the true radial distribution.

%Fig5
\begin{figure}
\resizebox{\hsize}{!}{\includegraphics{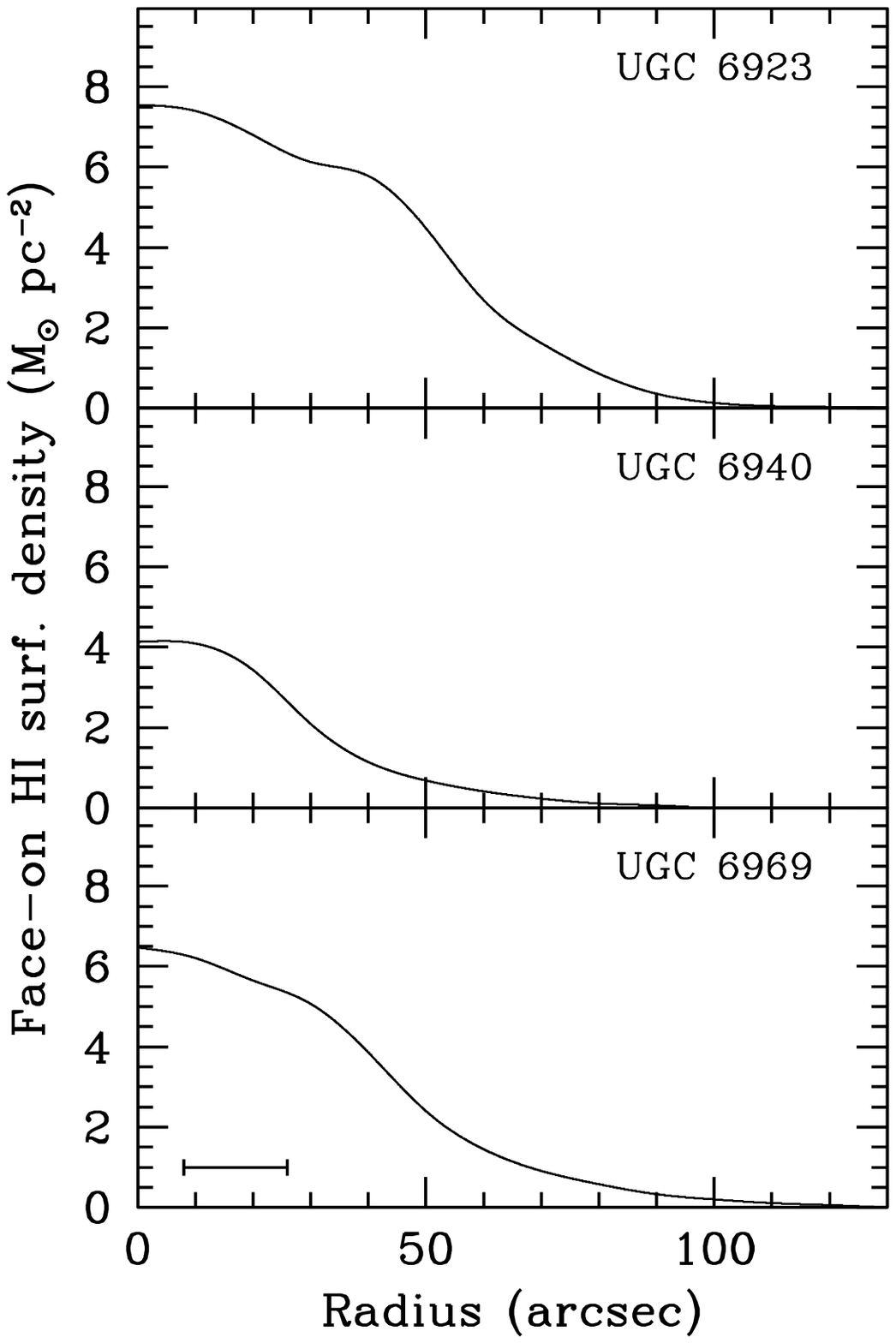}}
\caption[]{Deprojected \hi surface density as a function
of radius for the
three companions. The surface densities were 
obtained by averaging the total \hi maps
over elliptic annuli with the same orientations
as used for the rotation curve determination.
The approximate FWHM resolution is indicated by 
the errorbar.}
\end{figure}

\section{Construction of the velocity fields}

Inspection of the line profiles revealed that at certain
positions in the galaxy UGC 6923, these are rather skewed.
This skewness must be ascribed to beam smearing. If in one beam
there is a gradient in the true velocity field,
extensions will be created in the observed line profiles towards
the systemic velocity. It is difficult, if not impossible to
analytically correct for such beam smearing (see Begeman 1989).
A good approximation of the true radial velocity, however, is that
velocity at exactly the peak of the asymmetric profile. Of course,
this is only valid if the velocity field, gas distribution, and
resulting line profiles are regular. The problem is now to find
this true peak position.

In the past attempts have been made to do so by, for instance,
correcting the velocity side furthest from systemic for the instrumental
resolution and intrinsic dispersion (Sancisi \& Allen 1978). 
That method is rather dependent on the experience of the interpretator
and not generally applicable. Another method is fitting a double
Gaussian to the line profile and retaining only the one furthest
from the systemic velocity. The main problem associated with this method
is that observed profiles are rarely sampled by more than four
to five data points, while two Gaussians already have six free
parameters. Resulting fits may therefore be unphysical.
Rather one would like to fit a skewed analytical profile to the data
with only four free parameters. A suitable function is a Gauss-Hermite
polynomial expanded so as to include only one asymmetric ($h3$) term
(van der Marel \& Franx 1993):

\begin{eqnarray}
f(v) &=& a {\rm e}^{\frac{ {-(v-b)}^2}{2c^2} }
\biggl\{ 1 + h3 \biggl[ 1.1547 \left( \frac{v-b}{c} \right)^3
\nonumber \\
 &\;\;\; -& 1.1732 \left(\frac{v-b}{c} \right) \biggr] \biggr\},
\end{eqnarray}
where $v$ is the radial velocity and $a$, $b$, $c$, and $h3$ are
the parameters to be determined. In general $a$, $b$, and $c$ deviate
slightly from the Gaussian fit parameters if a pure Gaussian had been
fitted to the data. The true maximum of the profile is not at a velocity
equal to the value of $b$, but can easily be found numerically for
every case.

The steps in determining the velocity field are in principle the same
as those followed for the main galaxy. The initial estimates for the
fit were found by fitting a Gaussian to the conditionally transferred
data cube. With these estimates the skewed profile given above was
then fitted to the whole (full resolution) data cube. Line profiles
with dispersions less than 10 \kmss and amplitudes less than 1.5 times
the noise level were rejected. The position of the maximum was then
adopted as the value for the velocity field. The field was inspected
for continuity which led for all galaxies to the removal of typically
less than 10 pixels. Resulting velocity fields are displayed
in the bottom panels of Figs. 2, 3, and 4 for UGC 6923, UGC 6940,
and UGC 6969 respectively. The velocity field of UGC 6923 shows
an amount of differential rotation while the other two galaxies
have a velocity morphology which is close to solid body. As mentioned
before, the South East extension of UGC 6969 has a deviating velocity. 
Such a deviation should not be included in a rotation curve fit to the
velocity field an has therefore been removed from the velocity
field of UGC 6969 in Fig. 4. It is, however, still present in the
major axis x,v diagram in Fig. 9, top panel. Except for this feature
there are no appreciable irregularities present in the velocity fields.

As a bonus also the $h3$ image for the galaxies is generated.
For UGC 6923, the galaxy for which this asymmetric fitting was devised,
the $h3$ field is symmetric with respect to the centre. Values of
$h3$ are positive on one side along the major axis and negative on
the other side of the centre. This is a clear signature of beam
smearing. The other galaxies only show a very small and peculiar
asymmetry which will be discussed in the next section. The specific
beam smearing figure is not present. 

\section{The rotation curves}

A tilted ring model has been fitted to the velocity fields
of the companion galaxies to obtain the rotation curves.
Non overlapping rings have been used with a width of 10\arcsec\
and a weighting factor has been assigned to the data proportional
to the cosine of the angle measured from the major axis.
In first instance the dynamic position has been determined
using an initial estimate of the rotation velocity and 
orientation angles. 
Having this centre fixed, first the position angle
was determined, then the inclination, and after fixing that,
the rotation curve.

The application of this procedure to the three companions is not 
straightforward, which is caused by the close to solid body nature
of the velocity field. If one has the hypothetical case of a pure
solid body then the velocity contours are all straight lines
perpendicular to the major axis. By means of a tilted ring fit
it is then not possible to determine independently the dynamical 
position nor is it possible to obtain the inclination. In such a case
it has to be assumed that the optical position and inclination
coincide with the kinematic position and inclination. For \hi gas
within the optical region of a galaxy this is generally a valid assumption.
The three companions have velocity fields with an appreciable solid
body content. Therefore in some cases for some parameters optical
values have to be used, which will be a reasonable approximation
because the gas does not extend outside the optical image. 
For the outer tilted rings there will be positions along
the rings with no determined \hi radial velocities. To illustrate
this, the outermost ring for which a rotational value has been
determined is displayed as an ellipse on top of the velocity fields.
In that way it can be inspected which positions of the galaxy
are, and which are not, contributing to the rotation curve determination. 

The least squares fitting method gives errors, but these
are only formal errors, which are not always a good representation
of the true deviation from the data. To come
up with a more realistic error of the rotational velocity,
the fitting procedure has been repeated for the receding
and approaching side of the galaxy separately. Positions
and orientation angles were kept fixed at the same values
as for the whole galaxy and rotation velocities were
determined. The difference in rotation of the two sides gives
a better representation of the true error. The final error
is then given by the quadratic sum of the formal fit error
plus half the difference between the two sides. 

The rotation curve fitting procedure will now be discussed for
the three companions separately:

\subsection{UGC 6923}

There was no drifting of the dynamic centre as a function of radius,
which can therefore be determined quite accurately. The systemic velocity
is 1066 $\pm$ 2 \kmss and other fitted parameters can be found in
Table 3. Further steps in the fitting procedure are illustrated in
Fig. 6. In the top panel the position angle is shown when all three
parameters, rotation, inclination and pa were left free. The position
angle does not show a trend as a function of radius and was fixed
at a constant value of 343\degr. The middle panel of Fig. 6 gives
the fit result for the inclination after the pa is fixed.

%Fig6
\begin{figure}
\resizebox{\hsize}{!}{\includegraphics{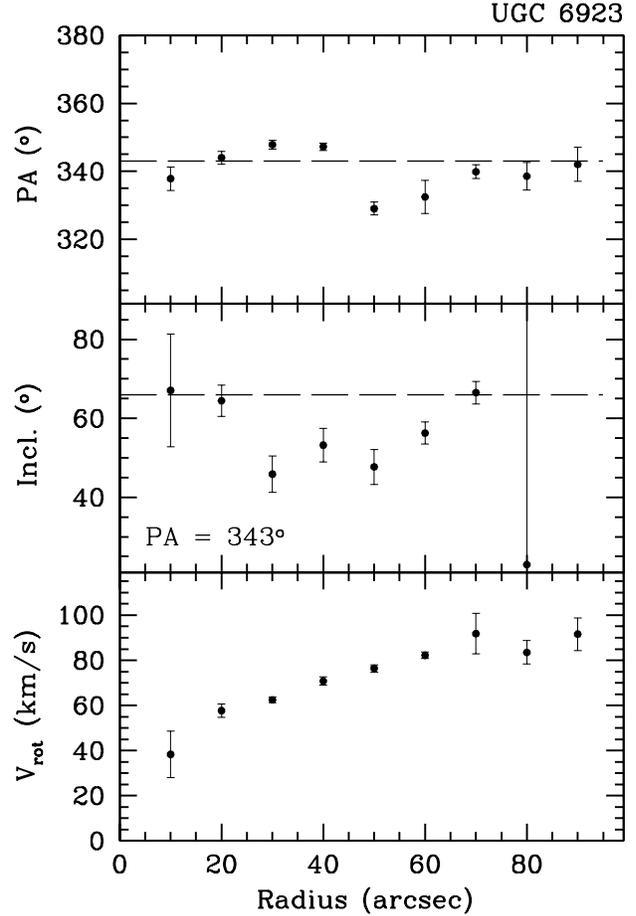}}
\caption[]{Determination of the rotation curve by
a tilted ring fit for UGC 6923.
In first instance the orientation angles and
rotation velocity were all left as free parameters
producing position angles as a function of 
radius (top panel). The PAs were fixed
at 343\degr\ indicated by the dashed line, and
the fitting procedure was rerun producing
the inclinations (middle panel). The optical value
of 66\degr\ was used for all inclinations
(dashed line). In the bottom panel
the final rotation curve is presented.}
\end{figure}

As noted above, it is difficult to get well determined, independent
kinematic inclination values. The optical inclination for UGC 6923
is 66\degr\ for an adopted $q_0$ of 0.11. This inclination is given
by the dashed line in Fig. 6 (middle panel), and is not inconsistent
with the kinematic inclination. Around a radius of
40\arcsec\ the kinematics prefers a somewhat smaller inclination
which is likely caused by the enhanced \hi density and accompanying
slightly deviating velocity field West of the centre. I feel confident that
the constant optical inclination is a good representation of the
actual gas kinematics. 

The lower panel
shows the final rotation curve for that fixed inclination. Errors
superposed on the points are already those when the difference between
the two sides has been taken into account.
All numerical values for the rotation curve of UGC 6923 are given
in Table 3. The outermost ring for which the rotation is determined
has a radius of 90\arcsec\ and is displayed as an ellipse in Fig. 2, 
bottom panel. This ring still picks up a number of radial velocity
points at diverse and independent positions of the galaxy. Therefore
the rotational value at that radius is a reliable and significant
measurement. 

\subsection{UGC 6940}

The velocity field of UGC 6940 is 
close to solid body and so small compared to the beam
that it was impossible to determine independently the dynamical centre.
Therefore it had to be assumed that the spatial position of the
optical nucleus coincides with the galaxy's dynamical centre. Using that
position and a guess for the rotation and orientation angles it
was possible to get a solid value for the systemic velocity of
1107 $\pm$ 2 \kms. All fit values for this galaxy are summarized in
Table 4. Next steps in the fitting procedure are illustrated in Fig. 7,
analogous to, but slightly different from Fig. 6. The middle panel shows
the fit for a free rotation, pa, and inclination, while the top panel
gives the fit for the position angle with the inclination
fixed at the optical value. The adopted constant values
for the position angle and inclination are equal to the optical
values and nicely compatible with the results of the tilted ring fit,
as can be seen by the dashed lines in Fig. 7. The rotation curve
plus the final errors are presented in the lower panel. 

%Fig7
\begin{figure}
\resizebox{\hsize}{!}{\includegraphics{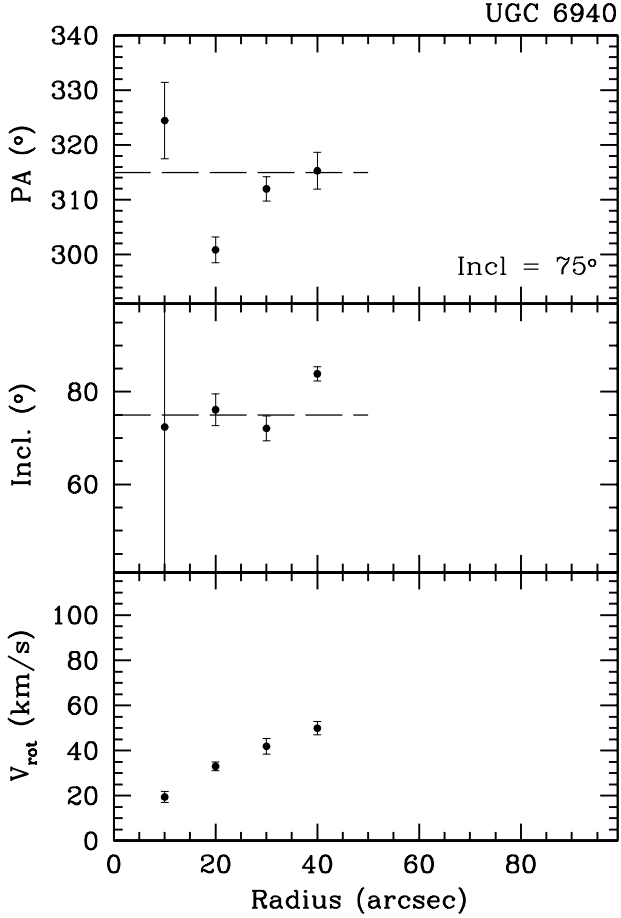}}
\caption[]{As Fig. 6, but now for UGC 6940.
For this galaxy the optical values for
both, the position angle and inclination
had to be adopted, being consistent with
the data points in top and middle panel.}
\end{figure}

\subsection{UGC 6969}

Using an initial estimate of the rotation together with fixed values
of the orientation angles resulted in a slight drift of the centre
as a function of radius of 20\arcsec\ in the spatial direction and
10 \kmss in velocity, over the 80\arcsec\ radial extent of the tilted
ring fit. To assess the reality of the drift, in a first iteration
the dynamical centre was fixed at the radially averaged value, and
a fit of the rotation was made. This rotation curve was used as a next
guess to redetermine the centre. Now the drift had disappeared
and the constant value of the dynamical centre is given in Table 5.
The optical centre was measured and was only 3\arcsec\ away from
the dynamical centre position, which is well within the measurement
errors. This demonstrates that the iteration has converged to
a reliable value, though for the nearly solid body velocity field
of UGC 6969 the procedure can never be done in a completely independent
manner. Having the centre fixed, the further fitting
process is illustrated in Fig. 8. There is no trend as a function
of radius for the fit of the orientation angles. Again the optical
values (for $q_0 = 0.11$) are fully compatible with the results
of the tilted ring fit, and were adopted. The rotation curve with
its proper errors is shown in the lower panel of Fig. 8.

%Fig8
\begin{figure}
\resizebox{\hsize}{!}{\includegraphics{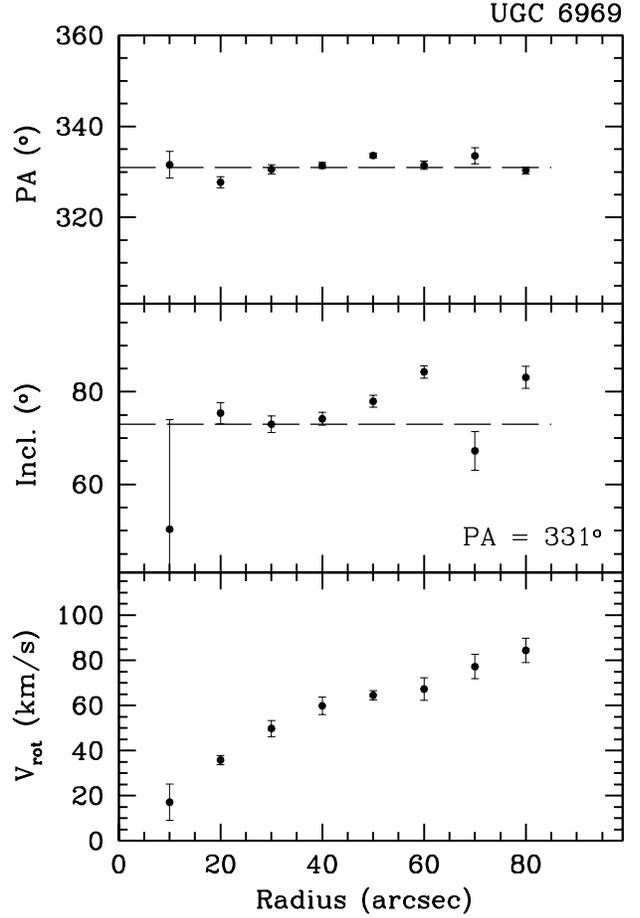}}
\caption[]{As Fig. 6, but now for UGC 6969.}
\end{figure}

%end subsubsections, do not know if there is a command to do so.
\subsection{Discussion}

To show the reliability and consistency of the method to derive
the rotation curves, in Fig. 9, the determined rotational
velocities have been
converted to radial velocities and plotted on the x,v diagrams
along the major axis. For all three galaxies the rotation points
are at the positions where one would expect them to be. NGC 6923,
for example, shows the skewed profiles in Fig. 9. The rotation curve
of this galaxy nicely follows the positions of maximum intensities
of the x,v diagram, demonstrating that the skew linefit procedure
is correct. 

In some cases, at the outer positions of the gas distributions
there is a rotational point given while there is no gas associated with
it in the x,v diagram. This is caused by the fact that the gas is spatially
not distributed in an elliptical shape similar to the orientation
of the kinematic tilted rings. Gas may then not be present along the
major axis, while there is still gas with associated radial velocities
at other positions along the tilted ring. For instance for UGC 6969
at the NW side of the galaxy, at that spot there is no \hi gas. Still
the outermost data point of the rotation curve is determined from radial
velocities of 22 pixels. Note that for this galaxy the cloud with
deviating velocity is displayed on the SE perimeter of the x,v diagram.
The affected region is not included in the tilted ring fit, but
still the rotation curve is fully consistent with the x,v diagram in the
sense that it correctly traces the envelope of the gas distribution 
(Sancisi \& Allen 1978).

%Fig9
\begin{figure}
\resizebox{7.2truecm}{!}{\includegraphics{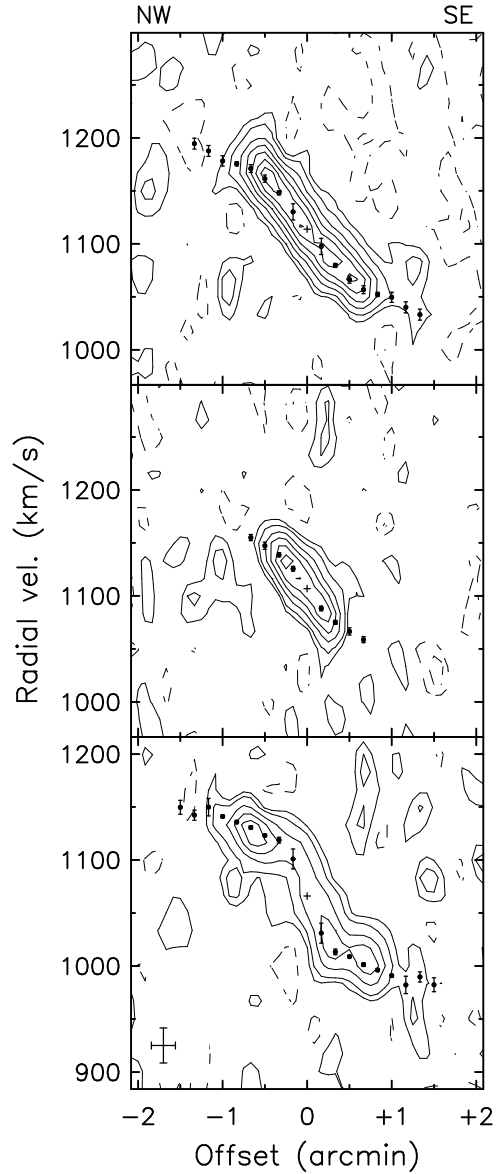}}
\caption[]{The rotation curves of the companions converted
to radial velocities, overplotted on a
full resolution position - velocity map along
the major axes. Top, middle, and bottom
panels are for UGC 6969, UGC 6940, and UGC 6923
respectively. The cross indicates the
position of the dynamic centre; contour
levels are at -3.92 and -1.96~K (dashed), and at
1.96, 3.92, 7.84, 11.8, 15.7, 19.6, and 23.5~K.
One can notice that for UGC 6923 (bottom panel)
a satisfactory correction for skewed line
profiles has been made.
}
\end{figure}

Though UGC 6923 and UGC 6969 have nearly the same maximum rotation
the shapes of the rotation curves are different. UGC 6969 has a rotation
curve which has more a solid body signature. Is this real?
Inspection of the x,v diagrams in Fig. 9 demonstrates that these
diagrams are indeed rather different for both galaxies. UGC 6923 shows
signs of beam smearing which can be observed in case of differential
rotation. UGC 6969, on the other hand, does not show these signs
and cannot show these signs if the galaxy has a rotation close to solid
body. In order for a differentially rotating galaxy to show a solid
body x,v diagram, the gas must have a central density depression in
combination with substantial beam smearing. As can be seen in Fig. 5,
where the observed radial density
profile is given, this is not the case. We must therefore conclude that
the different shapes of the two rotation curves is indeed real.

For the three companions Verheijen has also derived the rotational
parameters. His resolution is much worse, however, and he was forced
to use optical orientation values from the onset. This study confirms
that these optical values are appropriate and hence the global rotation
of Verheijen is equal to the present rotation. With the higher resolution
and better signal to noise it is now possible to derive rotation curves
with more details and going somewhat further out.

Looking at Fig. 9, one can at first notice that all companions rotate
in the same way with respect to the plane of the sky. Secondly, both
UGC 6969 and UGC 6940 exhibit a slight skewness of the line profiles
in the sense that the tail extends to higher velocities. This is visible
along the whole major axis and does not change significantly with position.
A possible explanation might be that we are witnessing the effect of
tidal forces when the two companions move past the large galaxy NGC 3992. 

\section{Decomposition of the rotation curves}

The decompositions of the rotation curves have, in principle, been done
in the same way as for the major galaxy. The I-band photometry 
of Tully et al. (1996) was used to represent
the radial mass profile of the galaxies (see
Fig. 10). Locally
isothermal stellar discs were assumed with a $z_0$ value of one fifth of the
scalelength. For the gas the radial distribution was taken equal to
that given in Fig. 5, multiplied with a factor of 1.4
to account for Helium, while the disc was infinitely
thin. Two kinds of dark halo have been considered, a pseudo 
isothermal one and an NFW halo.

The density distribution ${\rho}_{\rm h}$of a pseudo isothermal halo
is given by
\begin{equation}
{\rho}_{\rm h} = {\rho}_{\rm h}^0 \left[ 1 + \frac{R^2}{R^2_{\rm core}}
\right]^{-1},
\end{equation}
with a rotation law
\begin{equation}
v_{\rm h} = v_{\rm h}^{\rm max} \sqrt{
1 - \frac{R_{\rm core}}{R} \arctan \left( \frac{R}{R_{\rm core}}
\right) },
\end{equation}
where $R_{\rm core}$ is the core radius related to the maximum
rotation of the halo $v_{\rm h}^{\rm max}$ by
\begin{equation}
v_{\rm h}^{\rm max} = \sqrt{ 4\pi G {\rho}_{\rm h}^0
R^2_{\rm core} }.
\end{equation}
For an NFW halo the density distribution takes the form
\begin{equation}
{\rho}_{\rm NFW} = \frac{{\rho}_i}{(R/R_s)(1+R/R_s)^2},
\end{equation}
where $R_s$ is a characteristic radius and ${\rho}_i$ is related
to the density of the universe at the time of collapse.
The rotation curve $v_{\rm NFW}$ following from this distribution is
\begin{equation}
v_{\rm NFW} = 2.15\; v_{\rm max} \sqrt{ \frac{R_s}{R} {\rm ln}
\left( \frac{R}{R_s} + 1 \right) - \frac{R_s}{R+R_s} },
\end{equation}
where the maximum rotation $v_{\rm max}$ is reached at a 
radius of $2.16\; R_s$. When a certain cosmology is chosen the
structural parameter becomes related to the total mass of the halo.
In this case, there is a relation between the characteristic
radius $R_s$ and the maximum rotation. Presently I assume the
current concordance model for the cosmology: a low density CDM
universe with a flat geometry, called CDM$\Lambda$ with ${\Omega}_0
= 0.25,\; \Lambda = 0.75,$ and a Hubble constant of 75 \kmss per
Megaparsec. For that cosmology Navarro et al. (1997) relate the
parameter $M_{200}$ to $v_{\rm max}$ in their Fig.~7. Some manipulation
with equations leads to the following 
relation between $R_s$ and $v_{\rm max}$ 
\begin{equation}
\frac{R_s}{[{\rm kpc}]} = 0.0127 \left( \frac{v_{\rm max}}
{[{\rm km/s}]} \right)^{1.37},
\end{equation}
such that when Eqs. (6) and (7) are combined there is only
one free parameter left for the dark halo.

Besides the specific NFW profile, cosmological CDM simulations
also generate a certain scatter in these profiles.
Bullock et al. (2001) give in their Fig. 4 the scatter of the
concentration parameter $c$, which appears to be independent
of the halo mass. The positive $1\sigma$ deviation is approximately
0.105 on a $^{10}$log scale (factor 1.27) but the negative
$1\sigma$ deviation from the concentration parameter appears
to be larger, approximately 0.21 in $^{10}$log equal to a factor 1.62.
A different investigation by Jing (2000) gives for a $\Lambda$CDM
cosmology and for the most virialized halos a $1\sigma$ scatter
of 0.17 on a natural log (ln) scale (factor 1.19). For less 
virialized halos the scatter is a factor 1.28 and on average it is 1.22.
All these numbers differ somewhat, but a factor 1.28 or a 
$1\sigma$ deviation in $\Delta$ln$c$ of 0.25 seems a reasonable
compromise. Equation (7) has been derived from the $M_{200}$ versus
$v_{\rm max}$ relation and it is easy to demonstrate that a scatter in 
ln$(c)$ then translates in exactly the same scatter in ln$(R_s)$.
For a $3\sigma$ less concentrated dark halo, the coefficient in
Eq. (7) has to be multiplied by a factor e$^{3 \times 0.25}$, 
increasing it to 0.027. Such a $3\sigma$ deviation is considered
as a limit to the NFW functionality.

%Fig10
\begin{figure}
\resizebox{\hsize}{!}{\includegraphics{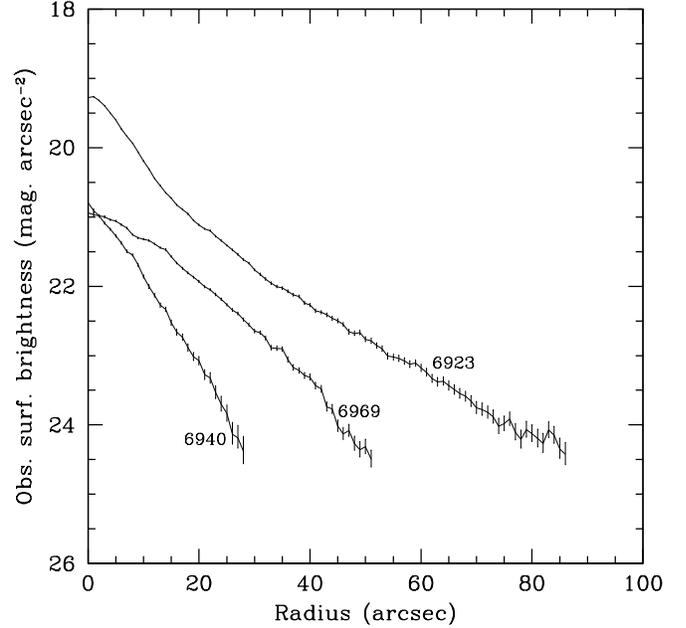}}
\caption[]{Radial profiles in the I-band for
the three companions as measured by Tully et al. (1996).}
\end{figure}

The results of the rotation curve decompositions are shown graphically
in Figs. 11, 12, and 13 for respectively UGC 6923, 6940, and 6969, while
Table 6 gives the numerical values. 
For an isothermal halo three cases have been investigated. A maximum disc
situation where the contribution of the stellar disc to the total
rotation is maximized, shown in the top panel of Figs. 11 to 13.
Also a minimum disc case is investigated, where there is simply no
stellar disc at all, of which the results are only given in Table 6.
Lastly, the decomposition is given for $(M/L)_I = 0.82$ in the middle
panels of the Figs. 11 and 13 and in the lower panel of 
Fig.~12. This situation corresponds to the 63\% criterion
for large galaxies and the ensuing M/L ratio. A derivation of this
mass-to-light ratio of 0.82 in the I-band is given in Sect.~6.

%Fig11
\begin{figure}
\resizebox{\hsize}{!}{\includegraphics{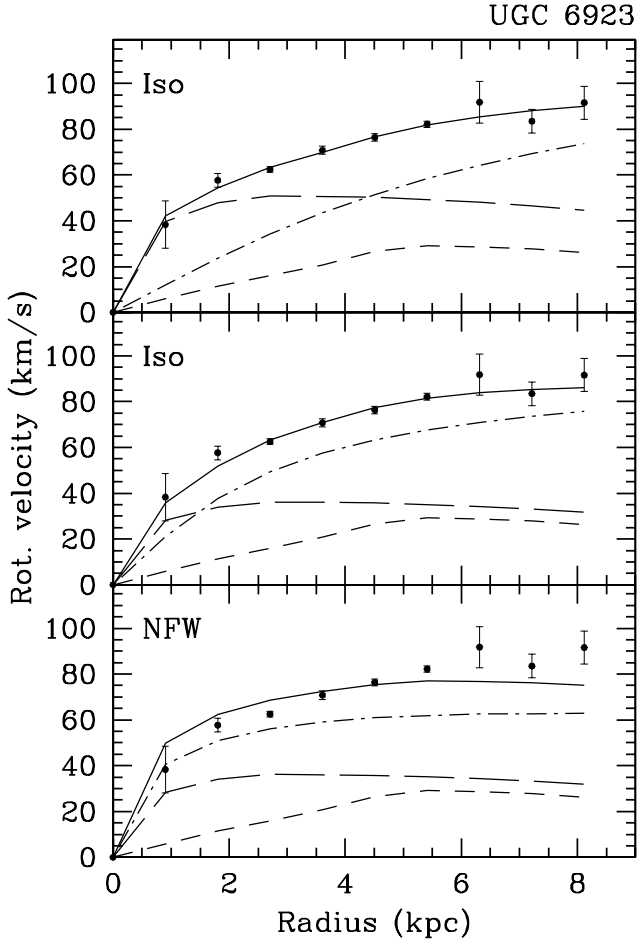}}
\caption[]{Rotation curve decompositions for UGC 6923.
The dots are the observed rotational
data and the full drawn line is the fit
to these. Individual contributions of
the disc (long dashed line), gas (short
dashed line), and dark halo (dash - dot line) are indicated.
Numerical values are given in Table~6.
{\it Top:} An isothermal halo maximum disc fit.
{\it Middle:} A decomposition for an isothermal halo and
$(M/L)^I_{\rm disc} = 0.82$.
{\it Bottom:} For an NFW-CDM$\Lambda$ halo with $(M/L)^I_{\rm disc} = 0.82$.
}
\end{figure}

%Fig12
\begin{figure}
\resizebox{\hsize}{!}{\includegraphics{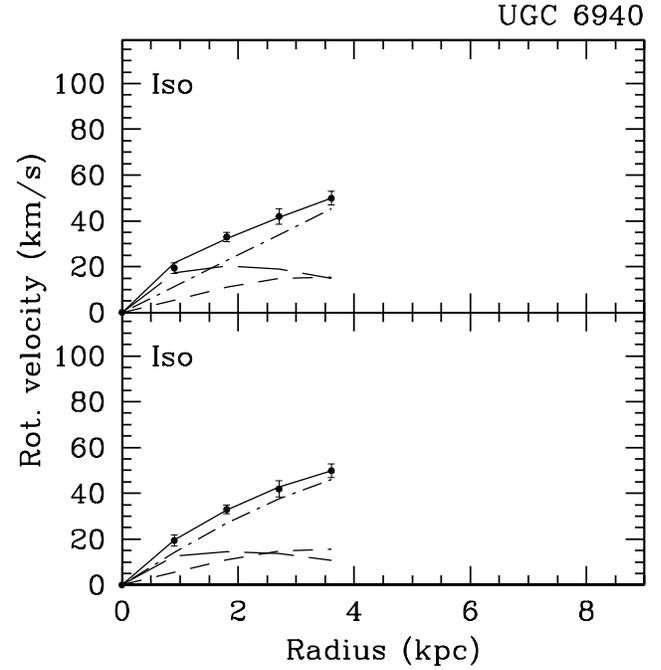}}
\caption[]{As Fig. 11, but now for UGC 6940. In this case no NFW fit
has been made.}
\end{figure}

%Fig13
\begin{figure}
\resizebox{\hsize}{!}{\includegraphics{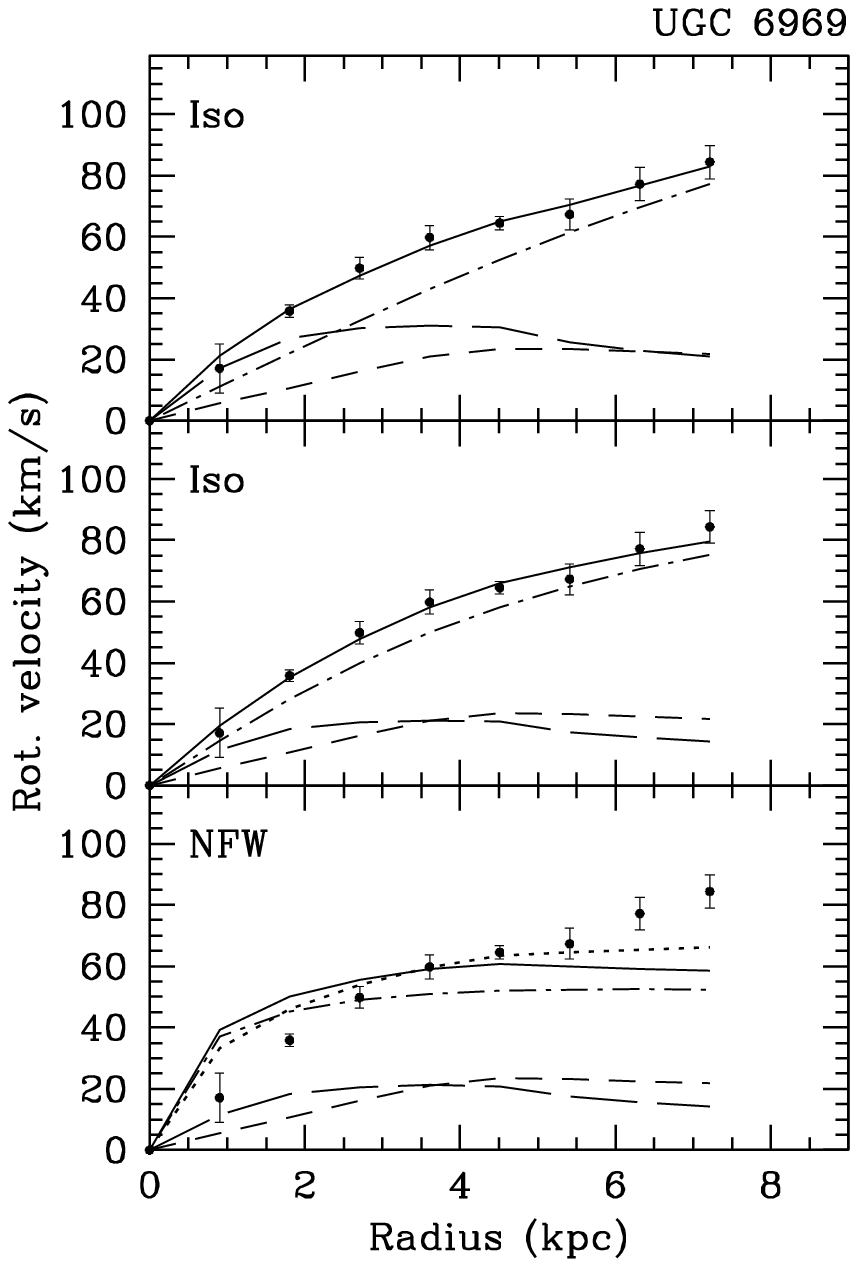}}
\caption[]{As Fig. 11, but now for UGC 6969.
In addition the dotted line in the lower panel gives the fit
of an NFW halo being less concentrated by $3\sigma$ from the
standard functionality.
}
\end{figure}

In case of maximum disc the $M/L_I$ ratio
of the dwarf galaxies ranges between 1.6 and 1.8 while
the core radius is comparable to or larger than the outer
measured radius of the gas. 
Reducing the stellar disc contribution
results in a smaller core radius of a few kpc. In all three cases the
fits to the observed data points are equally good and so, as usual,
from a rotation curve decomposition nothing can be deduced about
the contribution of the stellar disc. If, on the other hand, one assumes
$(M/L)_I = 0.82$ for all three galaxies and one compares the determined
dark halo rotations in Fig. 14 one notices that the dark halos of
the three companion galaxies are nearly equal. This is striking because
the total luminosities of the galaxies cover a range of a factor 16.
Do we have a situation where there are three identical dark halos
with different luminous galaxies embedded?

%Fig14
\begin{figure}
\resizebox{\hsize}{!}{\includegraphics{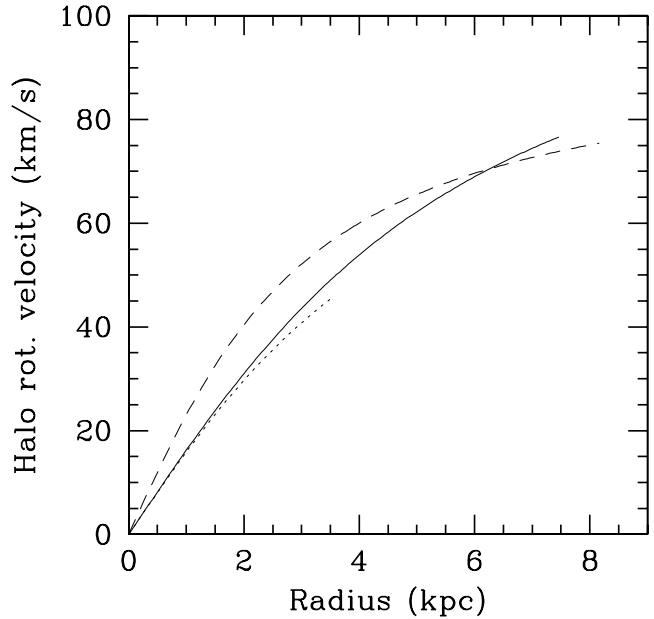}}
\caption[]{Rotation curves of the dark halos
of UGC 6923 (dashed), UGC 6940 (dotted),
and UGC 6969 (full drawn) in case $(M/L)^I_{\rm disc} = 0.82$.
Though these galaxies differ by more
than a factor of ten in luminosity, their
dark halo rotation is identical to
within the errors. This suggests that
different amounts of luminous matter have
settled in identical dark halos.}
\end{figure}

Let us leave out UGC 6940 because the halo rotation is not established
very far out. UGC 6923 and UGC 6969 differ by a factor five in luminosity,
while having approximately the same dark halo rotation. The effect of this
can be seen nicely in Figs. 11 and 13, middle panels. UGC 6969 has
the solid body rotation and the dark halo dominates over the whole
radial extent. UGC 6923 is more differentially rotating which is caused
by the larger disc mass contribution simply because the disc of UGC 6923
is more massive. For a $(M/L)_I$ of 0.82 the stellar disc still dominates
the rotation for radii less than 1.5 kpc. The differences between these
two galaxies have an interesting effect on their positions in the
Tully-Fisher relation, which will be discussed in the next section.

NFW decompositions have been made for UGC 6923 and UGC 6969 of which
the results in case of a fixed $(M/L)_I = 0.82$ are presented in 
Figs. 11 and 13, lower panels.
For UGC 6923 the fit of the strict NFW functionality (Eqs.~6$+$7) is
slightly off in the inner and outer parts. If the M/L ratio is reduced
to 0.0 or when a moderate deviation from the strict profile
within statistical limits is allowed, a good agreement between fit
and data points can be achieved. Consequently for UGC 6923 both the
isothermal and NFW halo can give a good description of the rotation. 

For UGC 6969, however, the situation is different. As can be seen
in Fig. 13, the strict NFW halo cannot fit the observations. 
A $3\sigma$ less concentrated halo has an $R_s$ parameter of 
8.4 kpc, which is nearly three times as large as for the strict case. 
Even then the fit is not good; the inner and outer rotational
points deviate too much from the total fitted rotation curve. 
Decreasing the M/L ratio of the disc barely improves matters because
the disc contribution is already small. If the $R_s$ and 
$v_{\rm max}$ parameters are left unrelated to each other,
the then ensuing two parameter fit diverges to very large values of $R_s$ and
$v_{\rm max}$. In such a case a small galaxy is embedded in
an unrealistically large dark halo. 
As discussed in Sect.~4, the x,v diagram of UGC 6969 in combination
with the derived radial density structure shows that there is no
apparent beam smearing present. 
In addition, beam smearing is expected to occur only in the most
edge-on or most unresolved galaxies (Swaters et al. 2000; McGaugh et
al. 2001) while for the more favourable objects, such as UGC 6969,
the derived \hi rotation curves generally give a good impression
of the actual rotation. 
The possibility remains of a slight
amount of beam smearing, but that can never account for the large
discrepancy between observations and NFW halo decomposition. 

One might wonder why for the isothermal case nearly equal dark halos
produce good fits to all the rotation curves and there are
problems with one galaxy for an NFW halo. A solution to this 
controversy can be found by inspection of Figs. 11 and 13. Like
for the isothermal situation also for the NFW case the fitted halo
rotation curves are very similar for UGC 6923 and UGC 6969. However,
an NFW halo has a cusp producing large rotations in the inner regions
which can be allowed for UGC 6923 but not for UGC 6969.

\section{Comparing the galaxies of the NGC 3992 group}

A detailed description now exists for the four galaxies in the NGC 3992
group, though the main uncertainty which remains is the actual
ratio of dark to luminous matter. A comparison will be made with
the other galaxies of the UMa cluster
which may generate additional information. The TF relation combines
in principle the kinematics, governed by dark and luminous matter,
with the total light. Consequently this relation and the position in this
relation of the four galaxies now studied should give insight into
the dark matter content. The I-band TF relation will be considered,
because one wants a band as red as possible to minimize dust and
population effects, while for the near infrared bands the total luminosities
of the galaxies are considered somewhat uncertain. Also the following
discussion on mass-to-light ratios will be conducted in the I-band.
In first instance one then needs the absorption free total light
of the four galaxies.

\subsection{Absorption corrections}

Various prescriptions exist for converting the observed light of
a galaxy into absorption free luminosities. For instance 
the method of Tully \& Fouqu\'e (1985), which does
a total conversion, while other methods are generally restricted to
a conversion to face-on luminosities. This method gives an 
internal absorption correction
$A^i_X$ for passband $X$ and inclination $i$ of

\begin{eqnarray}
A^i_X &=& -2.5\; {\rm log}10 \biggl[ f \left( 1 + {\rm e}^{-{\tau}{\rm sec}
(i)}\right) + (1 - 2f)
\nonumber \\
 &\;\;\; & \cdot \left( \frac{ 1 - {\rm e}^{-{\tau}{\rm sec}(i)}}
{{\tau}{\rm sec}(i)} \right)\biggr].
\end{eqnarray}
For an assumed $f$ parameter of 0.1 and $\tau$ values
of ${\tau}_B = 0.81, {\tau}_R = 0.40, {\tau}_I = 0.28,$ and
${\tau}_{\kprime} = 0.035$ one determines a correction for a face
on galaxy to absorption free $A_X^{i=0}$ of 0.40, 0.21, 0.15, and
0.02 for the B, R, I, and $\kprime$ band respectively.

The conversion to absorption free absolute magnitudes
$M_{T,X}^{b,i}$ in band $X$ is given for galaxies of the UMa cluster at
a distance of 18.6 Mpc by

\begin{equation}
M_{T,X}^{b,i} = m_T - A_X^b - A_X^i - 31.35\; ,
\end{equation}
where $m_T$ is the observed total brightness and $A_X^b$ is the
Galactic absorption correction which is generally very small. Both
these numbers can be found in Verheijen (1997).
For the galaxies
in the NGC 3992 group the observed total light and 
the internal absorption corrections according
to Eq.~(8) are given in Table~7.

There is, however, a vast difference in intrinsic luminosity between
NGC 3992 and the three surrounding dwarf galaxies. As demonstrated by
Tully et al. (1998) and by Giovanelli et al. (1997) there is a large
range in absorption corrections for galaxies with a range of intrinsic
luminosities. Bright, large galaxies have a large dust absorption
while small and certainly the low surface brightness galaxies have
generally a small or even negligible dust absorption. Consequently it
is unrealistic to use the Tully-Fouqu\'e equation for these four galaxies.
Tully et al. (1998) give an internal absorption correction to face-on
($A_X^{i-0}$, so no total correction) of

\begin{equation}
A_X^{i-0} = {\gamma}_X {\rm log}(\frac{a}{b}),
\end{equation}
where $\frac{a}{b}$ is the major/minor axis ratio of the galaxy and
${\gamma}_X$ a factor dependent on absolute luminosity corrected
for galactic, k-correction, and internal absorption $M_X^{b,k,i}$
given in the I-band by

\begin{equation}
{\gamma}_I = -0.20\; (16.9 + M_I^{b,k,i}).
\end{equation}
When the appropriate values for the four galaxies are substituted
in Eqs. (10) and (11) one has  $A_I^{i-0}$(3992) = 0.26 and $A_I^{i-0}$(U6923)
= 0.13, while for the other two galaxies the correction is zero.
Now still an internal dust correction for the face-on situation has
to be made. It will be clear that the whole procedure for internal
absorption correction is rather uncertain. Therefore at present,
taking into account the numbers given by Eq.~(8)
and the numbers given by the luminosity dependent correction
to face-on a total correction has been adopted of 0.52 magnitudes for NGC 3992
and a zero correction for the dwarfs.

\subsection{Mass-to-light ratios}

In his analysis of the mass distribution of the galaxies in the
UMa cluster Verheijen (1997) finds that in case of the 63\% criterion
the average M/L ratio in the $\kprime$ band of 
10 HSB galaxies (excluding NGC 3992)
amounts to 0.38 $M_{\sun}/L_{\sun}^{\kprime}$ with a scatter of 20\%.
For a maximum disc situation the average mass to light ratio
of the same 10 galaxies is 0.60, being a factor 1.6 larger.
Because in individual cases the calibration of the $\kprime$ photometry
may be somewhat uncertain this M/L ratio in the $\kprime$ band is
converted to the I-band. To that aim for all non-LSB galaxies of
Verheijens sample the K-I colour has been determined, corrected for
internal absorption according to Eq.~(8). The average of this,
$<{\kprime}-I>^i$, was -1.55 and is used to convert the average 
mass-to-light ratio
in the $\kprime$ to the I-band. In case of the 63\% criterion one
then has $\langle M/L_I {\rangle}^i = 0.82$ and 1.30 for the maximum disc
situation. In Paper~I a comparison has been made with $M/L_I$
ratios of a sample of galaxies for which an absorption to face-on
only has been made. To convert the average
$\langle M/L_I {\rangle}^i$ of the ten UMa cluster galaxies to a 
face-on correction only the same factor of 1.27 (0.26 mag.) as for
NGC 3992 has been adopted. Then for the maximum disc case results a
$\langle M/L_I {\rangle}^{i-0} = 1.65$ with a range from 0.7 to 2.2.

The three companions have a maximum disc $(M/L_I)^i$ value of 
1.7 $\pm$ $\sim 0.3$ (see Tables 6 and 7). When applying the total
absorption correction to NGC 3992, its maximum disc $(M/L_I)^i$
equals 2.91 $\pm$ 0.07 which is a factor of 1.7 larger than that of 
its companions. Is this consistent? To answer this, the colours
of the four galaxies have been used to compare the stellar
populations of the systems. The $(B-R)^i$ colours were derived
from the luminosities given by Tully et al. (1996) and Verheijen (1997).
An absorption correction for NGC 3992 was made using Eq. (8) resulting
in $(B-R)^i_{\rm N3992} = 1.21$ while no absorption corrections
for the companions were made giving a $(B-R)^i_{\rm comp}$ ranging
between 0.8 and 0.95. Then, according to Table~1 of Bell \& de Jong
(2001), population synthesis models predict for such a colour difference
a difference in  $(M/L_I)^i$ of a factor $1.6 \rightarrow 2.0$,
being very consistent with the observed value. Consequently the
difference in maximum disc $M/L$ values between NGC 3992 and its
companions can well be explained by a different stellar population
of the galaxies. There is a caveat, however. In the argumentation
it is assumed that the bluer colours of the companions are caused
solely by population effects, while in reality a likely lower 
metallicity may contribute as well in making the companions bluer. 

For the UMa cluster the  $(M/L_I)^i$ maximum disc value is 1.30
with a range of $0.54 \rightarrow 1.71$. A much larger value of
$(M/L_I)^i = 2.91$ or 2.60 if a bulge is allowed is found
for NGC 3992. Hence the maximum disc mass-to-light ratio of
NGC 3992 is 2.2 times (or 2.0 with bulge) as large as for the cluster
in which it should reside. This finding is similar as that for
the less well determined $\kprime$ band photometry where
Verheijen (1997) finds a factor of 2.9 difference. It is not
possible to explain the $M/L$ difference by population effects.
The average $\langle I - \kprime {\rangle}^i$ colour of 22 HSB
galaxies in the UMa cluster is 1.55 $\pm$ 0.14 while the 
$(I - \kprime )^i$ colour of NGC 3992 is 1.48, being absolutely
normal. For the $(B-I)^i$ colour one has 
$\langle B-I {\rangle}^i_{\rm UMa} = 1.35 \pm 0.28$
and $(B-I)^i_{\rm N3992} = 1.52$ which is also normal.
The difference in mass-to-light ratio thus has to be explained
by another mechanism. Let us first investigate what happens in
case of the 63\% criterion. Then the average $(M/L_I)^i$ for
the cluster galaxies is 0.82 with a range of $0.65 \rightarrow 1.06$.
For the best fitting rotation curve decomposition the maximum
disc velocity is slightly below 63\% and has an $(M/L_I)^i = 1.11
\pm 0.12$. Here the discrepancy between NGC 3992 and the rest of the
cluster is less; there is only a factor of 1.35 difference. However
when the disc $(M/L_I)^i$ of NGC 3992 is put at
the average cluster value of 0.82 the then following disc mass
of 54.3~10$^9$ \msol\ cannot give a reasonable rotation curve
decomposition any more.
In order to make equal $M/L$ ratios one might consider to put
the $(M/L_I)^i$ for NGC 3992 at 1.30 equal to the maximum disc
value of the other galaxies. That
then results in a situation where NGC 3992 is substantially sub maximum
while the other galaxies in the cluster are at maximum. Because
NGC 3992 is barred one might have expected just the opposite
and therefore this option is not likely. When the whole argumentation
given above is considered, the conclusion has to be reached
that the mass-to-light ratio of NGC 3992 is at least a factor
of 1.35 larger than that of the average value of the UMa cluster. 

This cannot be explained by a different stellar population. One then
has to invoke a different IMF for NGC 3992, containing more low
mass stars. But for that there is no obvious reason. The easy way
out is, of course, by putting NGC 3992 and its companions at a
larger distance and so behind the UMa cluster. To make the mass-to-light
ratio a factor 1.35 (= 0.33 mag.) smaller, the NGC 3992 group
needs to be 3.0 Mpc behind the cluster. For a factor 2.2
(= 0.86 mag.) lower $M/L$ ratio the group should be 9 Mpc behind
the cluster. Allowing a change in distance to NGC 3992 even makes
it possible to create a larger contribution of the disc to the total
rotation compared to the rest of the galaxies in the UMa cluster
and thus solving some problems possibly associated with its
barred morphology. 

\subsection{The Tully-Fisher relation}

In Fig. 15 the I-band TF relation of Verheijen is reproduced. The
maximum rotations of UGG 6923 and UGC 6969 have changed slightly
because of the present study. The new values for these two galaxies
have been put into the TF diagram of Fig. 15 and it should be noted
that the triangle for UGC 6923 could have been replaced with a
filled dot, meaning that the flat part of the rotation curve has
been reached. Also the data point for UGC 6940 has been
added to the diagram.

%Fig15
\begin{figure}
\resizebox{\hsize}{!}{\includegraphics{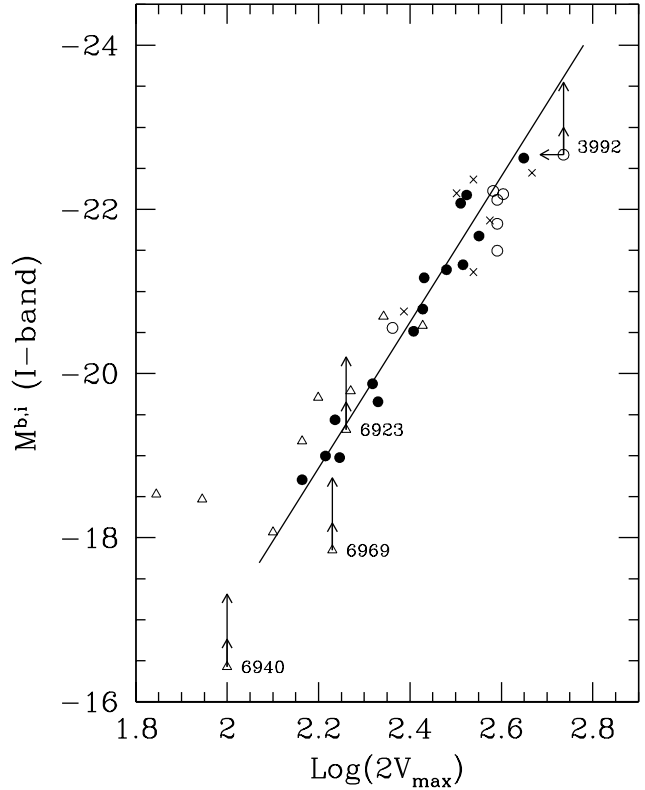}}
\caption[]{Reproduction of Verheijens (1997) Tully-Fisher
relation of the Ursa Major cluster. The
data point for UGC 6940 has been
added while the data points for UGC 6923 and UGC 6969
have been changed slightly according to the better determined
maximum rotations ($V_{\rm max}$) in the present paper.
The horizontal arrow attached to NGC 3992 indicates the shift
when instead of the maximum rotation
the flat part of the rotation curve is used.
Vertical arrows indicate the increase in luminosity with
0.33 or 0.86 mag. caused by a larger distance of 3 or 9 Mpc
respectively. The nearly equal rotation at different
luminosities for UGC 6923 and UGC 6969 can be
explained by different dark/luminous mass
ratios for these galaxies.}
\end{figure}

Let us first discuss the position of NGC 3992 in the TF relation.
It is obvious that NGC 3992 is too faint for its rotation, even
if instead of the maximum rotation that of the flat part is taken.
Doing the latter, NGC 3992 is 0.43 magnitude too faint. However, if
the light of the galaxy would be increased by more than a factor
1.35 ($>$ 0.33 mag.) the data point in the TF relation is consistent.
If the different mass-to-light ratio of NGC 3992 is explained by
a different IMF for that galaxy, only that data point in the TF 
relation has to be corrected. On the other hand, if a larger
distance is needed all the four data points belonging to the
NGC 3992 group in the relation (Fig. 15) have to be changed
by 0.33 mag. (factor 1.35) or by 0.86 mag. (factor 2.2). Considering
Fig. 15, such shifts are certainly possible. 

UGC 6923 and UGC 6969 have nearly the same maximum rotation, while
differing a factor of five in luminosity. Yet, assuming a $(M/L)_I^i$
of 0.82 the dark halos of these two galaxies are identical to within
the errors. This has an obvious effect on their position in the TF
relation as can be seen in Fig. 15. The argument can now be turned around,
the two galaxies have the same rotational velocity but differ
by a factor of five in brightness because they have a different
ratio of dark to luminous matter. A baryonic TF relation can be created
(McGaugh et al. 2000) by also taking the gas mass of a galaxy into account. 
In general the scatter of the TF relation then seems to decrease
for small galaxies. Using a fiducial gas $M/L$ ratio of 0.82,
the same as for the I-band rotation curve decomposition, the gas mass
of UGC 6923 and UGC 6969 was converted into light. The difference
in luminosity between the two galaxies then decreases
to a factor 2$\frac{1}{2}$, being still significant. 

Except for the usual problems
with absorption, populations effects, and inclination errors causing
observational or semi-observational scatter in the TF relation
there is also an intrinsic scatter, here twice illustrated by
the NGC 3992 group. At first, small differences in M/L (already corrected
for population effects and absorption) or in distance 
will cause data points to deviate.
Secondly, different ratios of luminous to dark mass will obviously
cause scatter. Matters are complicated because certainly for the smaller
galaxies the maximum of the rotation is not reached at the end
of the measured gas distribution. Considering all these problems
it is not surprising that the TF relation can be applied to
an individual case only with an appreciable error. For an ensemble of
galaxies spanning a wide range of luminosities, the relation
is a reflection of a globally equal ratio of dark to luminous matter.

\section{Discussion and conclusions}

This paper presents a nice example of two small galaxies which
appear to have nearly equal dark matter halos. Yet one of the galaxies
is considerably more luminous simply because it contains more
luminous matter which is dominant in the inner regions. The other
galaxy seems to be dominated by dark matter everywhere. As a hypothesis
one could imagine a situation where different amounts of gas
have fallen into the same halo, creating different amounts of stellar
material. In any case, it is inevitable that a different formation
scenario must have been at work, maybe associated with
an interaction with the dominant galaxy NGC 3992.

For UGC 6969 the universal NFW halo does not apply. In general a more
essential question is: what is the fraction of small galaxies
for which NFW is impossible? If that fraction appears to be small, 
those galaxies may have formed under exceptional circumstances
maybe involving large amounts of dissipational matter or
strong tidal interactions. The CDM structure formation is then
not in serious trouble. 

In my opinion the size of the abovementioned fraction is still
undetermined. This has to do with the inherent difficulty in
measuring rotation curves of LSB galaxies. The \hi observations
of de Blok \& McGaugh (1997) inevitably suffer from beam smearing
but as demonstrated by Swaters et al. (2000) beam smearing is most
associated with the smallest or close to edge-on galaxies.
For the more favourable objects of de Blok \& McGaugh the
derived rotation curves give a good impression of the actual rotation
even though the rotation curves were derived from the
x,v diagrams only. It is, of course, better to use the full
velocity field (as done in this paper) or to use H$\alpha$ observations
especially for the inner regions. Anyway, I expect these matters
can, and will be resolved in the near future. 

Finally a compilation of the main conclusions of this paper:

\begin{description}
\item[{\bf 1.}]
Detailed observations in the neutral hydrogen line have
been made of the large barred spiral galaxy NGC 3992
and its three small companion galaxies, UGC 6923, 
UGC 6940, and UGC 6969.
\item[{\bf 2.}]
In general the \hi distribution of all galaxies is regular.
NGC 3992 has a faint radial \hi extension outside its stellar disc;
contrary to
the companions where there is an abrupt end to the \hi gas distribution.
\item[{\bf 3.}]
For the three companions rotation curves have been derived
from their velocity fields.
\item[{\bf 4.}]
UGC 6923 and UGC 6969 have nearly the same maximum rotation.
Yet the shapes of the rotation curves are different which is likely
related to the factor of five difference in luminosity of 
the two.
\item[{\bf 5.}]
Assuming a reasonable M/L ratio for the three companions,
a decomposition of the rotation curves generates
nearly equal dark matter halos.
\item[{\bf 6.}]
An NFW-CDM$\Lambda$ dark halo is consistent with the observed
rotation curve of UGC 6923 but \underline{not} consistent with
the rotation curve of UGC 6969.
\item[{\bf 7.}]
A comparison has been made of the absorption corrected I-band
$M/L$ ratio of NGC 3992 and these ratios of other galaxies in the
UMa cluster. From that it can be concluded that or, the $M/L$ ratio
of NGC 3992 has to be at least 1.35 times larger than the average
ratio of the cluster galaxies, or, the NGC 3992 group is situated
more than 3 Mpc behind the cluster.
\item[{\bf 8.}]
This can also explain the position of NGC 3992 
in the Tully-Fisher relation
of the Ursa Major cluster, where NGC 3992 is approximately 0.43
magnitudes too faint for its rotation. 
\end{description}

\begin{acknowledgements}
The observations presented in this paper were obtained with
the Westerbork Synthesis Radio Telescope (WSRT) which is operated
by the Netherlands Foundation for Research in Astronomy (NFRA).
I thank Marc Verheijen for his help during the initial stages of
the data reduction and for intense discussions and useful remarks.
The Kapteyn Institute is acknowledged for providing hospitality and support.
\end{acknowledgements}

%\clearpage

%TABLE 1
\setcounter{table}{0}
\begin{table}
\caption[]{Observing parameters}
\begin{flushleft}
\begin{tabular}{ll}
\noalign{\smallskip}
\hline
\noalign{\smallskip}
Telescope & WSRT \\
Observing date & May 1997 to Sept. 1997 \\
Duration of observation & 4 $\times$ 12 h. \\
Number of interferometers & $\sim$ 27 \\
Baselines (min-max-incr.) & 36 - 2736 - 36 m. \\
Full res. beam (FWHM, $\alpha \times \delta$) & 14\arcsec $\times$ 18\arcsec \\
FWHpower primary beam & 37\arcmin \\
Rms (1$\sigma$) noise per channel & \\ 
\quad full res. & 1.96 K = 0.473 $M_{\sun}$pc$^{-2}$ \\
\quad res. = 30\arcsec $\times$ 30\arcsec & 0.55 K = 0.132 $M_{\sun}$pc$^{-2}$ \\
Velocity central channel & 1050 \kms \\
Bandwidth & 5 MHz \\
Number of channels & 64 \\
Channel separation & 16.6 \kms \\
Velocity resolution & 33.3 \kms \\
Field centre (1950) & (11$^{\rm h}$ 55$^{\rm m}$ 07$^{\rm s}$ ; 53\degr 39\arcmin 18\arcsec) \\
K-mJy conversion, & \\
\quad equivalent of 1 mJy/beam & 2.62 K (full res.) \\
 & 0.73 K (res. = 30\arcsec) \\
Adopted distance & 18.6 Mpc \\
\noalign{\smallskip}
\hline
\end{tabular}
\end{flushleft}
\end{table}

%TABLE 2
\setcounter{table}{1}
\begin{table}
\caption[]{Galaxy parameters}
\begin{flushleft}
\begin{tabular}{lll}
\noalign{\smallskip}
\hline
\noalign{\smallskip}
\quad NGC 3992 & & \\
Hubble type & SBb(rs)I & a \\
Brightness (in B) & 10.86 mag. & b \\
Brightness (in I) & 8.94 mag. & b \\
Opt. incl. ($q_0 = 0.11$) & 57\degr & b \\
Opt. PA major axis & 68\degr (= 248\degr) & b \\
PA major axis bar & 37\degr & c \\
Deprojected bar length & 150\arcsec & d \\
Scalelength & undef & \\
Total \hi mass & 5.9 10$^9$ \msol & d \\
21 cm cont. flux & 43.2 mJy & d \\
\noalign{\medskip}
\quad UGC 6923 & & \\
Brightness (in B) & 13.91 mag. & b \\
Brightness (in I) & 12.36 mag. & b \\
Opt. incl. ($q_0 = 0.11$) & 66\degr & b \\
Opt. PA major axis & 354\degr & b \\
Scalelength (in I) & {20}\farcs{9} & b \\
Total \hi mass & 0.64 10$^9$ \msol & e \\
21 cm cont. flux & $<$ 2.6 mJy & b \\
\noalign{\medskip}
\quad UGC 6940 & & \\
Brightness (in B) & 16.45 mag. & b \\
Brightness (in I) & 15.44 mag. & b \\
Opt. incl. ($q_0 = 0.11$) & 75\degr & b \\
Opt. PA major axis & 135\degr & b \\
Scalelength (in I) & {8}\farcs{52} & b \\
Total \hi mass & 0.16 10$^9$ \msol & e \\
21 cm cont. flux & $<$ 1.3 mJy & b \\
\noalign{\medskip}
\quad UGC 6969 & & \\
Brightness (in B) & 15.12 mag. & b \\
Brightness (in I) & 14.04 mag. & b \\
Opt. incl. ($q_0 = 0.11$) & 73\degr & b \\
Opt. PA major axis & 330\degr & b \\
Scalelength (in I) & {11}\farcs{65} & b \\
Total \hi mass & 0.44 10$^9$ \msol & e \\
21 cm cont. flux & $<$ 3.8 mJy & b \\
\noalign{\smallskip}
\hline
\multicolumn{3}{l}{a Sandage \& Tammann (1981)} \\
\multicolumn{3}{l}{b Verheijen (1997)} \\
\multicolumn{3}{l}{c Measured from photograph} \\
\multicolumn{3}{l}{d Paper I} \\
\multicolumn{3}{l}{e This paper} \\
\end{tabular}
\end{flushleft}
\end{table}

%TABLE 3
\setcounter{table}{2}
\begin{table}
\caption[]{The rotation curve of UGC 6923}
\begin{flushleft}
\begin{tabular}{llllll}
\noalign{\smallskip}\hline\noalign{\smallskip}
R & $V_{\rm rot}$ & ${\varepsilon}_{\rm vrot}$ &
R & $V_{\rm rot}$ & ${\varepsilon}_{\rm vrot}$ \\
(\arcsec) & (\kms) & (\kms) & (\arcsec) & (\kms) & (\kms) \\
\noalign{\smallskip}\hline\noalign{\smallskip}
10 & 38.3 & 10.3 & 60 & 82.2 & 1.4 \\
20 & 57.7 & 3.0  & 70 & 91.8 & 9.0 \\
30 & 62.5 & 1.2  & 80 & 83.5 & 5.2 \\
40 & 70.8 & 1.8  & 90 & 91.6 & 7.2 \\
50 & 76.4 & 1.6  &    &      &     \\
\noalign{\smallskip}\hline\noalign{\smallskip}
\multicolumn{6}{l}{Pos. of dynamical centre}\\
\multicolumn{3}{l}{\quad RA (1950)}&\multicolumn{3}{l}{
11$^{\rm h}$ 54$^{\rm m}$ {14}\fs{2} }\\
\multicolumn{3}{l}{\quad Declination (1950)}&\multicolumn{3}{l}{
53\degr 26\arcmin {20}\farcs{7}  }\\
\multicolumn{3}{l}{\quad $V_{\rm sys}$ (Hel.)}&\multicolumn{3}{l}{
1066 $\pm$ 2 \kms}\\
\multicolumn{3}{l}{Inclination}&\multicolumn{3}{l}{66\degr \enskip (Opt.)}\\
\multicolumn{3}{l}{P.A.}&\multicolumn{3}{l}{
343\degr $\pm$ 4\degr}\\
\noalign{\smallskip}\hline
\end{tabular}
\end{flushleft}
\end{table}

%TABLE 4
\setcounter{table}{3}
\begin{table}
\caption[]{The rotation curve of UGC 6940}
\begin{flushleft}
\begin{tabular}{llllll}
\noalign{\smallskip}\hline\noalign{\smallskip}
R & $V_{\rm rot}$ & ${\varepsilon}_{\rm vrot}$ &
R & $V_{\rm rot}$ & ${\varepsilon}_{\rm vrot}$ \\
(\arcsec) & (\kms) & (\kms) & (\arcsec) & (\kms) & (\kms) \\
\noalign{\smallskip}\hline\noalign{\smallskip}
10 & 19.4 & 2.4 & 30 & 41.9 & 3.5 \\
20 & 33.0 & 2.0 & 40 & 49.9 & 3.0 \\
\noalign{\smallskip}\hline\noalign{\smallskip}
\multicolumn{6}{l}{Pos. of dynamical centre}\\
\multicolumn{3}{l}{\quad RA (1950)}&\multicolumn{3}{l}{
11$^{\rm h}$ 55$^{\rm m}$ {12}\fs{6} }\\
\multicolumn{3}{l}{\quad Declination (1950)}&\multicolumn{3}{l}{
53\degr 30\arcmin {45}\farcs{3}  }\\
\multicolumn{3}{l}{\quad $V_{\rm sys}$ (Hel.)}&\multicolumn{3}{l}{
1107 $\pm$ 2 \kms}\\
\multicolumn{3}{l}{Inclination}&\multicolumn{3}{l}{75\degr \enskip (Opt.)}\\
\multicolumn{3}{l}{P.A.}&\multicolumn{3}{l}{
315\degr \enskip (Opt.)}\\
\noalign{\smallskip}\hline
\end{tabular}
\end{flushleft}
\end{table}

%TABLE 5
\setcounter{table}{4}
\begin{table}
\caption[]{The rotation curve of UGC 6969}
\begin{flushleft}
\begin{tabular}{llllll}
\noalign{\smallskip}\hline\noalign{\smallskip}
R & $V_{\rm rot}$ & ${\varepsilon}_{\rm vrot}$ &
R & $V_{\rm rot}$ & ${\varepsilon}_{\rm vrot}$ \\
(\arcsec) & (\kms) & (\kms) & (\arcsec) & (\kms) & (\kms) \\
\noalign{\smallskip}\hline\noalign{\smallskip}
10 & 17.1 & 8 & 50 & 64.5 & 2.1 \\
20 & 35.8 & 2 & 60 & 67.3 & 5.0 \\
30 & 49.8 & 3.6 & 70 & 77.2 & 5.4 \\
40 & 59.8 & 3.9 & 80 & 84.4 & 5.4 \\
\noalign{\smallskip}\hline\noalign{\smallskip}
\multicolumn{6}{l}{Pos. of dynamical centre}\\
\multicolumn{3}{l}{\quad RA (1950)}&\multicolumn{3}{l}{
11$^{\rm h}$ 56$^{\rm m}$ {12}\fs{9} }\\
\multicolumn{3}{l}{\quad Declination (1950)}&\multicolumn{3}{l}{
53\degr 42\arcmin {8}\farcs{6}  }\\
\multicolumn{3}{l}{\quad $V_{\rm sys}$ (Hel.)}&\multicolumn{3}{l}{
1114 $\pm$ 2 \kms}\\
\multicolumn{3}{l}{Inclination}&\multicolumn{3}{l}{73\degr \enskip (Opt.)}\\
\multicolumn{3}{l}{P.A.}&\multicolumn{3}{l}{
331\degr $\pm$ 2\degr}\\
\noalign{\smallskip}\hline
\end{tabular}
\end{flushleft}
\end{table}
\clearpage

%TABLE 6
\setcounter{table}{5}
\begin{table*}
\caption[]{Rotation decompositions of the companions}
\begin{flushleft}
\begin{tabular}{lllll}
\noalign{\smallskip}\hline\noalign{\smallskip}
UGC & & 6923 & 6940 & 6969 \\
\noalign{\smallskip}\hline\noalign{\smallskip}
Incl. (\degr) & & 66 & 75 & 73 \\
$h_I$ (kpc) & & 1.88 & 0.77 & 1.05 \\
Obs. light (10$^9$ $L^I_{\sun}$) & & 1.76 & 0.104 & 0.376 \\
Gas mass, \hi + He (10$^9$ \msol) & & 0.89 & 0.22 & 0.622 \\
$(M/L)_I^{\rm obs}$ max disc & Iso & 1.64 $\pm$ 0.23 & 1.6 $\pm$ 0.2 & 1.78 $\pm$ 0.4 \\
$\rcore$ max disc (kpc) & Iso & 5.3 $\pm$ 2.2 & $>$ 20 & 9.2 $\pm$ 5.2 \\
$\vhmax$ max disc (\kms) & Iso & 124 $\pm$ 57 & $>$ 500 & 199 $\pm$ 114 \\
$\rcore$ min disc (kpc) & Iso & 1.39 $\pm$ 0.16 & 2.0 $\pm$ 0.2 & 2.93 $\pm$ 0.33 \\
$\vhmax$ min disc (\kms) & Iso & 93 $\pm$ 13 & 73 $\pm$ 8 & 104 $\pm$ 13 \\
$\rcore$ $(M/L)^i_I = 0.82$ (kpc) & Iso & 2.24 $\pm$ 0.27 & 3.45 $\pm$ 0.4 & 4.24 $\pm$ 0.6 \\
$\vhmax$ $(M/L)^i_I = 0.82$ (\kms) & Iso & 94 $\pm$ 13 & 97 $\pm$ 11 & 121 $\pm$ 19 \\
$R_s$ $(M/L)^i_I = 0.82$ (kpc) & NFW & 3.7 $\pm$ 0.2 & & 2.9 $\pm$ 0.5 \\
$v_{\rm max}$ $(M/L)^i_I = 0.82$ (\kms) & NFW & 63 $\pm$ 3 & & 53 $\pm$ 6 \\
$R_s$ $(M/L)^i_I = 0$ (kpc) & NFW & 4.8 $\pm$ 0.15 & & 3.25 $\pm$ 0.45 \\
$v_{\rm max}$ $(M/L)^i_I = 0$ (\kms) & NFW & 77 $\pm$ 2 & & 57 $\pm$ 6 \\
$R_s$ $(M/L)^i_I = 0.82$ (kpc) & NFW-$3\sigma$ & & & 8.4 $\pm$ 1.3 \\
$v_{\rm max}$ $(M/L)^i_I = 0.82$ (\kms) & NFW-$3\sigma$ & & & 66 $\pm$ 7 \\
\noalign{\smallskip}\hline
\end{tabular}
\end{flushleft}
\end{table*}

%TABLE 7
\setcounter{table}{6}
\begin{table*}
\caption[]{Light and mass-to-light ratios}
\begin{flushleft}
\begin{tabular}{llllllllll}
\noalign{\smallskip}\hline\noalign{\smallskip}
Galaxy & $L_I^{\rm obs}$ & Incl. & $A_I^i$ & $A_I^i$ & $L_I^i$ & 
$(M/L)_I^i$ & Mass for & Mass & Mass \\
 & & & (Eq. 8) & used & & max disc & $(M/L)_I^i = 0.82$ & max disc &
best fit \\
 & (10$^9$ $L_{\sun}$) & (\degr) & (mag.) & (mag.) & (10$^9$ $L_{\sun}$) &
({\msol}/$L_{\sun}$) & (10$^9$ \msol) & (10$^9$ \msol) & (10$^9$ \msol) \\
\noalign{\smallskip}\hline\noalign{\smallskip}
N3992 & 41.16 & 57 & 0.26 & 0.52 & 66.27 & 2.91 & 54.3 & 194.1 & 73.7 \\
U6923 & 1.76  & 66 & 0.34 & 0    & 1.76  & 1.64  & 1.44  & 2.89  & - \\
U6940 & 0.104 & 75 & 0.52 & 0    & 0.104 & 1.6 & 0.085  & 0.17  & - \\
U6969 & 0.376 & 73 & 0.47 & 0    & 0.376 & 1.78 & 0.308  & 0.67  & - \\
\noalign{\smallskip}\hline
\end{tabular}
\end{flushleft}
\end{table*}
\clearpage


\begin{thebibliography}{}
\bibitem[1987]{}
Begeman K. 1987, Ph. D. Thesis, University of Groningen\par
\bibitem[]{}
Begeman K. 1989, A\&A, 223, 47\par
\bibitem[]{}
Bell E.F., \& de Jong R.S. 2001, ApJ, 550, 212\par
\bibitem[]{}
Bosma A 1978, Ph. D. Thesis, University of Groningen\par
\bibitem[]{}
Bottema R. 1993, A\&A, 275, 16\par
\bibitem[]{}
Bottema R. 1997, A\&A, 328, 517\par
\bibitem[]{}
Bottema R., \& Verheijen M.A.W. 2002, to appear in A\&A (Paper I)\par
\bibitem[]{}
Bullock J.S., Kolatt T.S., Sigad Y., et al. 2001, MNRAS, 321, 559\par
\bibitem[]{}
Carignan C., \& Freeman K.C. 1985, ApJ, 294, 494\par
\bibitem[]{} 
Courteau S., \& Rix H.W. 1999, ApJ, 513, 561\par
\bibitem[]{}
de Blok W.J.G., \& McGaugh S.S. 1997, MNRAS, 290, 533\par
\bibitem[]{}
de Blok W.J.G., McGaugh S.S., Bosma A., \& Rubin V.C. 2001, ApJ, 552, L23\par
\bibitem[]{}
Freedman W.F., Madore B.F., Gibson B.K., et al. 2001, ApJ, 553, 47\par
\bibitem[]{}
Giovanelli R., Haynes M.P., Herter T., et al. 1997, AJ, 113, 22\par
\bibitem[]{}
Gottesman S.T., Ball R., Hunter J.H., \& Huntley J.M. 1984, ApJ, 286, 471\par
\bibitem[]{}
Jing Y.P. 2000, ApJ, 535, 30\par
\bibitem[]{}
Kalnajs A.J. 1983, in IAU Symp. 100, The internal kinematics
of galaxies, ed. Athanassoula, E., (Reidel, Dordrecht), p87\par
\bibitem[]{}
Kent S.M. 1986, AJ, 91, 1301\par
\bibitem[]{}
McGaugh S.S., Schombert J.M., Bothun G.D., \& de Blok W.J.G. 2000,
ApJ, 533, L99\par
\bibitem[]{}
McGaugh S.S., de Blok W.J.G., \& Rubin V.C. 2001, AJ, 122, 2381\par
\bibitem[]{}
Navarro J.F. 1998, astro-ph/9807084\par
\bibitem[]{}
Navarro J.F., Frenk C.S., \& White S.D.M. 1996, ApJ, 462, 563\par
\bibitem[]{}
Navarro J.F., Frenk C.S., \& White S.D.M. 1997, ApJ, 490, 493\par
\bibitem[]{}
Pickering T.E., Impey C.D., van Gorkom J.H., \& Bothun G.D. 1997,
AJ, 114, 1858\par
\bibitem[]{}
Sakai S., Mould J.R., Hughes S.M.G., et al. 2000, ApJ, 529, 698\par
\bibitem[]{}
Salucci P., Ashman K.M., \& Persic M. 1991, ApJ, 379, 89\par
\bibitem[]{}
Sancisi R., \& Allen R.J. 1978, A\&A, 74, 73\par
\bibitem[]{}
Sandage A., \& Tammann G.A. 1981, A Revised Shapley-Ames Catalog
of Bright Galaxies, Carnegie Institute of Washington\par
\bibitem[]{}
Sellwood J.A., \& Moore E.M. 1999, ApJ, 510, 125\par
\bibitem[]{}
Swaters R.A., Madore B.F., \& Trewhella M. 2000, ApJ, 531, L107\par
\bibitem[]{}
Tully R.B., \& Fisher J.R. 1977, A\&{A}, 54, 661\par
\bibitem[]{}
Tully R.B., \& Fouqu\'e P. 1985, ApJS, 58, 67\par
\bibitem[]{}
Tully R.B., \& Pierce M.J. 2000, ApJ, 533, 744\par
\bibitem[]{}
Tully R.B., \& Verheijen M.A.W. 1997, ApJ, 484, 145\par
\bibitem[]{}
Tully R.B., Verheijen M.A.W., Pierce M.J., et al. 1996, AJ, 112, 2471\par
\bibitem[]{}
Tully R.B., Pierce M.J., Huang J-.S., et al. 1998, AJ, 115, 2264\par
\bibitem[]{}
van Albada T.S., Bahcall J.N., Begeman K., \& Sancisi R. 1985, ApJ, 295, 305\par
\bibitem[]{}
van Albada T.S., \& Sancisi R. 1986, Phil. Trans. R. Soc. London, Ser. A,
320, 447\par
\bibitem[]{}
van der Marel R.P., \& Franx M. 1993, ApJ, 407, 525\par
\bibitem[]{}
Verheijen M.A.W. 1997, Ph. D. Thesis, University of Groningen\par
\bibitem[]{}
Verheijen M.A.W., \& Sancisi R. 2001, A\&A, 370, 765\par
\end{thebibliography}
\end{document}